\documentclass[twocolumn]{svjour3}         
\smartqed  
\usepackage{graphicx}
\usepackage{mathptmx}      
\usepackage{natbib}
\journalname{SSRv}
\begin{document}

\title{Obtaining Spectra of Turbulent Velocity from Observations
}

\newcommand{\be}{\begin{equation}}
\newcommand{\ee}{\end{equation}}
\newcommand{\ba}{\begin{eqnarray}}
\newcommand{\ea}{\end{eqnarray}}
\newcommand{\eps}{\varepsilon}


\author{A. Lazarian}


\institute{
              Department of Astronomy, University of Wisconsin-Madison \\
              \email{lazarian@astro.wisc.edu}           
}


\maketitle

\begin{abstract}
We discuss a long-standing problem of how turbulence can be studied using observations of Doppler broadened 
emission and absorption lines. The focus of the present review is on two new techniques, the Velocity-Channel
Analysis (VCA), which makes use of the channel maps, and the Velocity Coordinate Spectrum
(VCS), which utilizes the fluctuations measured along the velocity axis of the  Position-Position
Velocity (PPV) data cubes. Both techniques have solid theoretical foundations based on analytical 
calculations as well as on numerical testings. 
Among the two the VCS, which has been developed quite recently,
has two unique features. First of all, it is applicable to turbulent volumes that
are not spatially resolved. Second, it can be used with absorption lines that
do not provide good spatial sampling of different lags over the image of the turbulent object.
In fact, recent studies show that measurements of absorption line along less than 10 absorption directions are sufficient
for a reliable recovering of the underlying spectrum of the turbulence. Moreover, both weak
absorption lines and absorption lines in saturated regime can be used, which extends the
applicability of the technique.
Our comparison of the VCA and the VCS with a more traditional technique of Velocity Centroids
shows that the former two techniques recover reliably the spectra of supersonic turbulence,
while the Velocity Centroids may be used only for studying subsonic turbulence.
We discuss spectra of astrophysical turbulence obtained with the VCA and the VCS techniques.
\end{abstract}

\section{Introduction}
\label{intro}

The ISM is turbulent on scales ranging from AUs to kpc (see Armstrong
et al 1995, Elmegreen \& Scalo 2004), with an embedded magnetic field
that influences almost all of its properties. MHD turbulence is accepted to be
of key importance for fundamental astrophysical processes, e.g.
 star formation, propagation and acceleration of cosmic rays. 

How to study astrophysical turbulence? This review stresses the importance of
observational studies of the turbulence spectra. We feel that the progress of numerical
modeling of astrophysical turbulence shifted somewhat the attention of the astrophysical
community from observational studies. Therefore we believe that stressing of the
synergy of the observational and numerical studies is due. 

It is generally accepted that numerical simulations have tremendously influenced our understanding of the physical
conditions of turbulence (see Vazquez-Semadeni et al. 2000, Mac Low \& Klessen 2004,
Ballesteros-Paredes et al. 2007, McKee \& Ostriker 2007 and ref. therein). Present codes can
produce simulations that resemble observations, but because of their limited numerical resolution, there are serious concerns
about how well they reproduce reality, especially  astrophysical turbulence (see e.g. McKee 1999, Shu et
al. 2004).  In this respect, observational studies of turbulence can test to what extent the numerical approach represents the
actual physical processes. 

The turbulence spectrum, which is a statistical measure of turbulence,
 can be used to compare observations with both numerical simulations and theoretical
 predictions. Note that statistical descriptions are nearly indispensable strategy when
dealing with turbulence. The big advantage of statistical techniques
is that they extract underlying regularities of the flow and reject
incidental details. The Kolmogorov description of unmagnetized incompressible
turbulence is a statistical one. For instance it predicts that the difference in velocities at
different points in turbulent fluid increases on average
with the separation between points as a cubic root of the separation,
i.e. $|\delta v| \sim l^{1/3}$. In terms of direction-averaged
energy spectrum this gives the famous Kolmogorov
scaling $E(k)\sim 4\pi k^2 P({\bf k})\sim k^{5/3}$, where $P({\bf k})$ 
is a {\it 3D} energy spectrum defined as the Fourier transform of the
correlation function of velocity fluctuations $\xi ({\bf r})=\langle  
\delta v({\bf x})\delta v({\bf x}+{\bf r})\rangle$. Note that in
this paper we use $\langle  ...\rangle$ to denote averaging procedure.

However, the importance of obtaining turbulence spectrum from observations goes
well beyond testing the accuracy of numerical modeling. 
 For instance, the energy spectrum 
$E(k)dk$ characterizes how much
energy resides at the interval of scales $k, k+dk$. At large scales $l$
which correspond to small wavenumbers $k$ ( i.e. $l\sim 1/k$) one expects
to observe features reflecting energy injection. At small scales
one should see the scales corresponding to
sinks of energy. In general, the shape of the spectrum is
determined by a complex process of non-linear energy transfer and
dissipation. Thus, observational studies of the turbulence spectrum can determine
sinks and sources of astrophysical turbulence. 

In view of the above it is not surprising that attempts
to obtain spectra of interstellar turbulence have been numerous since 1950s
(see Munch 1958). However, various directions
of research achieved various degree of success (see 
Armstrong, Rickett \& Spangler 1995). 
For instance, studies of turbulence statistics of ionized media 
accompanied by theoretical advancements in understanding scattering and scintillations of radio waves in
ionized medium (see Goodman \& Narayan 1985) were rather successful
(see Spangler \& Gwinn 1990). They provided the information of
the statistics of plasma density at scales $10^{8}$-$10^{15}$~cm. 
However, these sort of measurements provide only the density statistics, 
which is an indirect measure of turbulence. 

Velocity statistics is a much more coveted turbulence measure. 
Although it is clear that Doppler broadened lines
are affected by turbulence, recovering of velocity statistics is
extremely challenging without an adequate theoretical insight.
Indeed, both $z$-component velocity
and density contribute to fluctuations of the energy density $\rho_s ({\bf X}, V_z)$ in the 
Position-Position-Velocity (PPV) space.

Traditionally, the information on turbulence spectra is obtained with the measure of Doppler shift termed Velocity Centroids,
$\sim \int V_z \rho_s dV_z$, where the integration is taking place over the range of the velocities relevant to the object under study.
In this situation it is easy to see that the Velocity Centroids are also proportional to $\int V_z \rho ds$, where $\rho$ is an
actual three dimensional density and the integration is performed along the line of sight (see Lazarian \& Esquivel 2003).
 While usually the Velocity Centroids are normalized over the integrated intensity over the line of
sight (see Stenholm 1990), Esquivel \& Lazarian (2005) showed that this normalization does not change the statistical properties of the measure.
The numerical and analytical analysis in Esqivel \& Lazarian (2005) and Esquivel et al. (2007) showed that the Velocity Centroids fail
for studying supersonic turbulence. This provides extremely bad news for the studies of velocity statistics in molecular clouds and diffuse cold ISM (see Dickman \& Kleiner 1985, Miesch et al. 1999, Miville-Deschines et al. 2003). The studies for HII regions (O'Dell et al. 1987) are less affected, as in most cases, the turbulence there is subsonic. 

There were attempts to analyze PPV data cubes in other ways, for instance, Crovisier \& Dickey (1983), Green (1993) and
Stanimirovic et al. (1999) analyzed power spectra of velocity channels of HI data. The spatial spectrum of fluctuations of these velocity slices of PPV revealed power-law dependences, but the physical meaning of these dependences was absolutely unclear. 

The analytical study of the statistical properties of the PPV energy density $\rho_s$ has been initiated in Lazarian \& Pogosyan (2000). There the observed statistics of $\rho_s$ was related to the underlying 3D spectra of velocity and density in the astrophysical turbulent volume. Initially, the volume was considered transparent, but later the treatment was generalized for the volume with self-absorption and for studies of turbulence using absorption lines.  
In what follows, we discuss how the observable Doppler-shifted lines can be used to recover the spectrum of
turbulent velocity using two new techniques that, unlike other mostly empirical
techniques, have solid theoretical foundations. How to obtain using spectroscopic observations other characteristics of turbulence,
e.g. higher order statistics, anisotropies has been reviewed earlier (see Lazarian 2004).

The new techniques of studying astrophysical turbulence are intended to make use
of extensive spectroscopic surveys. In the review, we mostly discuss HI and CO data. However,
many more lines, not necessarily radio lines, can be used. For instance, absorption optical and UV lines
can also be used within the framework of the new techniques. 

\begin{figure*}
\hbox{
  \includegraphics[width=.3\textheight]{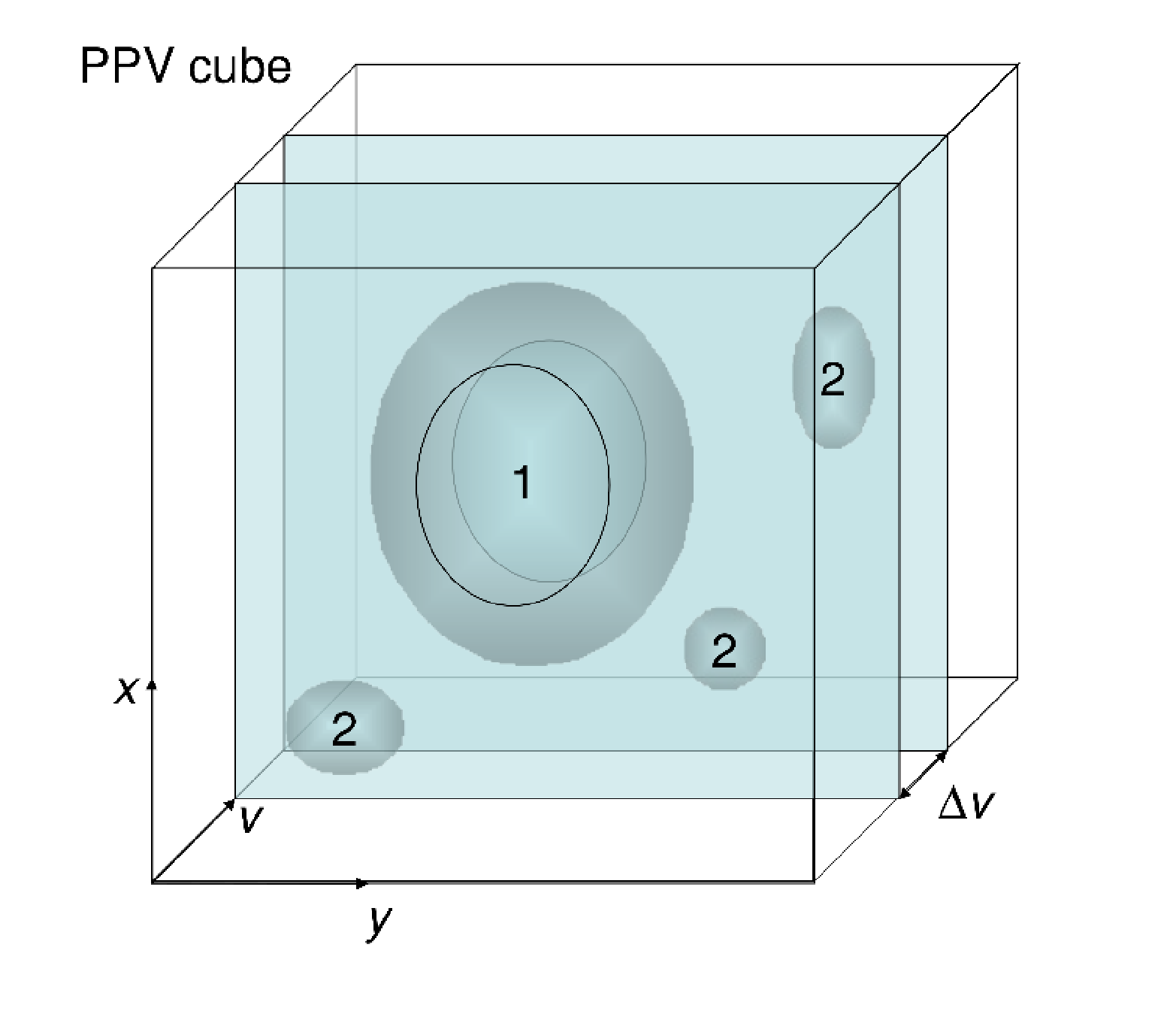}
\hfill
 \includegraphics[height=.3\textheight]{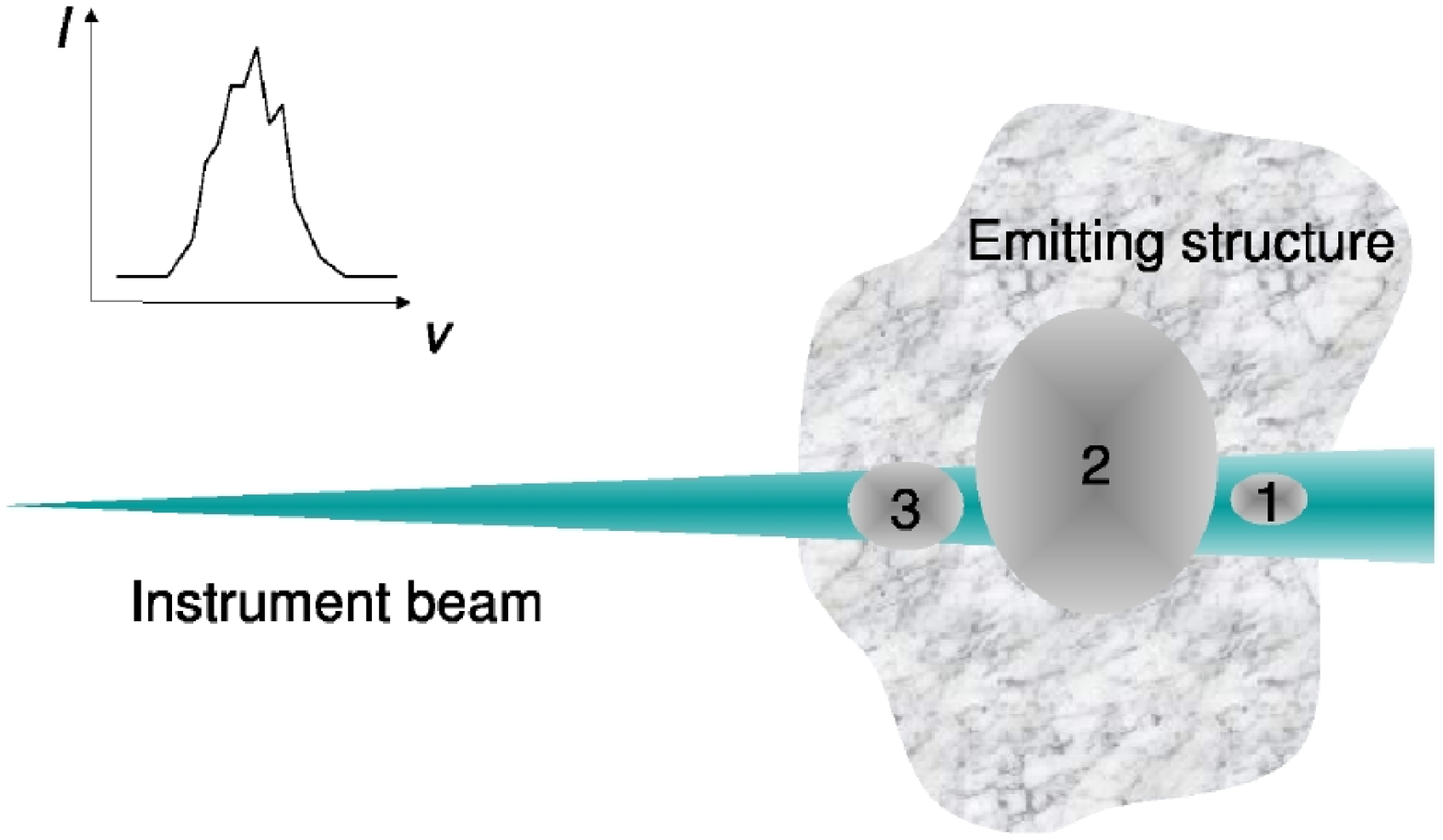}
}
  \caption{{\it Left Panel}: PPV data cube. Illustration of the concepts of the thick and thin velocity
slices. The slices are thin for the PPV images of the large eddies, but thick for the images of small
eddies. {\it Right Panel}: Illustration of the VCS technique. For a given instrument resolution large
eddies are in the high resolution limit, while small eddies are in the low resolution limit. The velocity profile
shown illustrates the turbulence-broderned profiles of lines that are used for the VCS studies.}
\label{fig1}
\end{figure*}

In general, this review attempts to provide an intuitive approach to the understanding of the new techniques of
data analysis. We start with a general discussion of how the Doppler-broaderned spectral lines can be analyzed in
\S 2, introduce the basics of the analytical description of the statistics of the spectral lines in \S 3. In \S 4 for
power-law spectra of velocities and densities we present the spectra of one, two and three dimensional
fluctuations observable spectral line fluctuations and discuss the effects of additional factors, e.g. finite resolution
of a telescope, presence of self-absorption. We describe the studies of turbulence with saturated absorption lines
in \$ 5, compare the two major techniques for studying turbulence with spectral lines in \S 6, and present the
results of numerical testing of the techniques in \S 7. In \S 8 we discuss the ways of extending the techniques, e.g.
to allow for a) studies of turbulence that does not obey a simple power-law, b) studies of turbulence in the local ISM,
and c) studies of turbulence with lines the strength of which is proportional to density squared. A short summary of
the observational studies of interstellar turbulence using the new techniques is presented in \S 9. We analyze the
alternative approaches to studying velocity turbulence in \S 10. In \S 11 we discuss the synergy that can be obtained by
combining different techniques of observational turbulence studies and outline the prospects for the field.  

\section{Ways to Analyze Position-Position-Velocity Data Cubes}
\label{sec:first}

The most detailed information on turbulent velocity observationally available from an emitting or absorbing turbulent volume is a Position-Position-Velocity (PPV) data cube. Evidently, the resolution in the  P-P plane is provided by the spatial resolution of the telescope, while the V-resolution requires spectroscopic resolution of the Doppler shifted lines at different locations over the image of the emitting volume. 

As we mentioned above, frequently, astronomers analyze the distribution of PPV intensities within a particular velocity range or a channel map. This velocity range may be the minimal interval corresponding to the maximum of spectral resolution (see Green 1993) or chosen to be wider, e.g. to decrease the noise (Stanimirovic et al. 1999).    
{\it Spatial} spectra obtained by taking Fourier transform of channel maps had been used to study HI 
before we conducted our theoretical study of what those spectra mean in 
Lazarian \& Pogosyan (2000, henceforth LP00). These studies were providing power spectra of  channel map
intensity distributions which varied from study to study. The relation of the power spectra to the underlying velocity fluctuations was a subject of speculation with claims that the spectral index of channel maps coincides with the spectral index of velocity fluctuation in 2D slices of the actual turbulent  volume (see Green 1990 and references therein). We showed in LP00 that the latter naive claim, as well as other speculations related to the interpretation of the spatial intensity variations within channel maps, are not correct.  This was done through establishing the
explicit relation between the statistics of the intensity fluctuations within the channel maps and the statistics of the underlying velocity and density. 

To understand the essence the Velocity Channel Analysis (VCA) formulated in LP00, consider a PPV cube 
arising from measuring Doppler shifted spectra from a turbulent volume (see Figure~\ref{fig1}). The channel maps in Figure~\ref{fig1} correspond to the velocity slices of the PPV cube. One may ask a question whether fluctuations  depend on the thickness of the velocity slice. It is intuitively clear that if the medium is optically thin and the 
velocity is integrated over the entire spectral line, the fluctuations can depend only on density inhomogeneities. It is also suggestive that the contribution of the velocity fluctuations may depend on whether the images of the eddies under study fit within a velocity slice or if their velocity extent is larger than the slice thickness (see Figure~\ref{fig1}, left). In the former case the slice is "thick" for eddies and in the latter case it is "thin".
According to LP00 the spectra of fluctuations that correspond to ``thin'' and ``thick'' slices are different and varying the thickness (i.e. effective channel width) of slices it is possible to disentangle the statistics of underlying velocities and densities
in the turbulent volume.  

Note, that the questions of whether the spectra of intensity fluctuations within channel maps may depend on the
channel map thickness were not posed by the research prior to LP00 study. This resulted in comparing of spectra
obtained with the channel map spectra of different thickness, which, as we understand now,
is  incorrect.  LP00 showed that some of the differences of the earlier reported
spectral indexes were due the differences in the thickness of the velocity slices analyzed (see Stanimirovic \& Lazarian 2001). By now the VCA technique has been applied to a number of different sets of spectral line data to
determine the underlying velocity and density of turbulent fluctuations (see \S 9).  

LP00 dealt with optically thin data. Optically thick CO data was traditionally analyzed differently from HI (see Falgarone \& Puget 1995). Namely, spectra of total intensities were studied. The origin of such a spectrum and its relation to the underlying velocity and density fluctuations was established in 
Lazarian \& Pogosyan (2004, henceforth LP04).
This work also clarified the effects of absorption that were reported for HI data. 
In terms of the techniques of turbulence study, LP04 extends the study of the VCA technique in LP00 to the case of studying turbulence within an emitting turbulent  volume in the presence of absorption. Intuitively, results of LP04 are easy to understand.
In turbulence there is a unique statistical relation between the physical scales and the turbulent velocities. LP04 proved that if the
thickness of a velocity slice is larger than the dispersion of velocities of the eddies which get optically thick, the effects of self-absorption should be taken into account (see \S 4 for more details). 

A radically different way of analyzing spectroscopic data is presented in Lazarian \& Pogosyan (2006, henceforth LP06). There the spectra of intensity fluctuations along the V-axis of the PPV cube are studied (see Figure~\ref{fig1}, right). The mathematical foundations of the technique can be traced to LP00, but there the high potential of the Velocity
Coordinate Spectrum (VCS), as the technique was later termed in Lazarian (2004), required further studies. Indeed, it took some time to understand the advantages that the VCS provides for the practical handling of the observational data. Numerical testing
of the technique (see Chepurnov \& Lazarian 2008) was also important.

LP06 deals with studies of turbulence using emission lines and absorption lines in the limit of weak absorption.  VCS allows us to analyze turbulence when spatial information is either not available or sparse. For instance, as we discuss in \S 6 one
can potentially study turbulence and obtain the turbulence spectrum in a spatially unresolved turbulent volume\footnote{Note, that poor resolution in terms of the PPV cube is equivalent to averaging over P-P dimensions.}. One can also study turbulence having just a couple of absorption lines, which corresponds to sampling of the PPV volume along a few directions only (see Figure~\ref{abs1}). Naturally, this stems from the fact that the fluctuations along the V-axis are studied by the VCS. This makes the VCS technique really unique for the velocity turbulence studies.
We illustrate the technique discussing its first applications in \S 9. 

A more recent study by Lazarian \& Pogosyan (2008, henceforth LP08) deals with the studies of turbulence using saturated absorption lines. There we showed that the saturation of the line acts as a sort of window function in the velocity space. In the presence of this window function one can still use the unsaturated wings of the line to get a high frequency input on turbulence.

\begin{figure}
  \includegraphics[width=.5\textwidth]{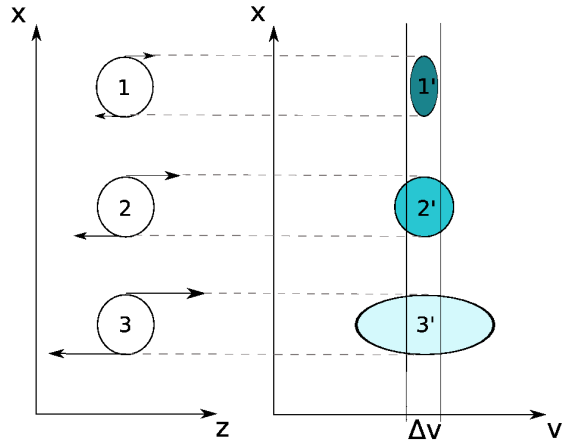}

  \caption{An illustration of the mapping from the real space to the PPV space. In the real space 3 eddies above have the same
  spatial size, but different velocities. They are being mapped to the PPV space and there they have the same PP dimensions, but a different
  V-size. The larger is the velocity of eddies, the larger the V-extent of the eddies, the less density of atoms over the image of the eddy. 
  Therefore, in terms of the intensity of fluctuations in the velocity channel $\Delta v$, the largest contribution is coming from the least energetic eddy, i.e.
  eddy 1,  while the most energetic eddy, i.e. eddy 3, contributes the least. }
\label{eddies}
\end{figure}

The non-trivial nature of the statistics of the eddies in the PPV space is illustrated in Figure \ref{eddies}. The figure illustrates the fact that from 3 equal size and equal density eddies, the one with the smallest
velocity provides the largest contribution to the PPV intensity. This explains the assymptotical scalings of power spectra in Table~3, which indicates that
  a spectrum of eddies that corresponds to most of turbulent energy at {\it large} scales corresponds to the spectrum of thin channel map intensity fluctuations having most of the energy at {\it small} scales. It is also clear that if the channel map or velocity slice of PPV data gets thicker than the
  velocity extent of the eddy 3, all the eddies contribute to the intensity fluctuations the same way, i.e. in proportion to the total number of
  atoms within the eddies. Similarly, in terms of the spectrum of fluctuations along the V-axis, the weak velocity eddy 1 provides the most singular small-scale contribution, which is important in terms of the VCS analysis.

\section{Basics of the Formalism}
While the mathematical foundations that relate the statistics of
the turbulent Doppler-shifted lines with the underlying turbulent spectra are provided in
mathematically-intensive papers, e.g., LP00, LP04, LP06, LP08. Below 
we provide a brief introduction to this formalism and introduce the mathematical foundations of the VCA and the VCS
techniques (see more in
LP00, LP04 and LP06).
Our goal is to relate the statistics
that can be obtained through spectral line observations, for instance,
the structure function of the intensity of emission $I_{\bf X}(v)$ 
\begin{equation}
{\cal D}({\bf X}, v_1,v_2)\equiv \left\langle \left[ I_{\bf X}(v_1) -
I_{\bf X}(v_2)\right]^2\right\rangle~~~,
\label{dv_emiss}
\end{equation}
where the $z$-axis component velocity
$v$ is measured in the direction defined by the two dimensional
vector\footnote{Henceforth we denote by the capital bold letters
the two dimensional position-position vectors that specify the line of sight.
Small bold letters are reserved to describe the vectors
of three dimensional spatial position.  The $z$-axis 
is chosen to be along the line of sight.}
${\bf X}$, to the underlying properties of the turbulent cascade.

Consider, for simplicity the case of no absorption. In this case one
can easily see that the density of energy in the PPV space 
\begin{equation}
\rho_s ({\bf X}, v)\sim \int^S_{-S} dz \rho({\bf X}, z) \phi_{vz}({\bf X}, z)
\label{first}
\end{equation}
where the emission is coming from the cloud of the $2S$ and the intensity
of emission is assumed to be proportional to the density $\rho$. The distribution
function of the z-component of velocity is given by $\phi_{vz}$.

What is the distribution function $\phi_vz$? It is a Maxwellian shifted by the z-components of
the turbulent $u$ velocity
\begin{equation}
\phi_{vz}({\bf x}) {\mathrm d} v =\frac{1}{(2\pi \beta)^{1/2}}
\exp\left[-\frac{(v-u({\bf x}))^2}
{2 \beta }\right] {\mathrm d} v ~~~,
\label{phi}
\end{equation}
where $\beta=\kappa_B T /m$, $m$ being the mass of atoms. For $T\rightarrow 0$ the
function $\phi_v$ tends to a delta-function that depends on regular gas flow and
the turbulent velocity $u$. 

Several things are clear from Eq.~(\ref{first}). First of all, both densities and velocities contribute to the
energy density in the PPV space. Velocity and density enter the expression in different ways. Thus
one can expect that the expressions for the statistics of $\rho_s$ will depend differently on the statistics of
$\rho$ and $u$.

What are the statistics that we are dealing with? We discuss the correlation and structure functions and
their Fourier components, which are the spectra. 
For the $z$-component of the turbulent velocity field
(i.e. $u$), we use the structure function 
\begin{equation}
D_z({\bf r})=
\langle(u({\bf x}+{\bf r})-u({\bf x}))^2 \rangle ~~, 
\label{Dz}
\end{equation}
which for a self-similar power-law turbulent motions
 provide 
 \begin{equation}
 D_z\sim D(L) (r/L)^m~~,
 \label{rm}
 \end{equation}
where $L$ is the turbulent injection scale, $D_z(L)$ is the variance of velocity at this scale,
$m$ is the scaling exponent, which is $m=1/3$ for the Kolmogorov turbulence.
These velocity correlations together with the correlations of over-density
\be
\xi(r)=\xi({\bf r}) = \langle \rho ({\bf x}) \rho ({\bf x}+{\bf r}) \rangle~~
~,
\label{xifirst}
\ee
enter the correlation function that can be constructed from the PPV densities $\rho_s$, which are
available through spectroscopic observations.
 If the gas is confined in an isolated cloud of size $S$, the zero-temperature correlation
function is (see LP06)
\begin{eqnarray}
\xi_s(R,v)\equiv \langle \rho_s({\bf X_1},v_1)\rho({\bf X_2}, v_2)\rangle\\ \nonumber 
\propto
\int_{-S}^S {\mathrm d}z \left(1-\frac{|z|}{S}\right)
\; \frac{\xi( r)}{D_z^{1/2}({\bf r})}
\exp\left[-\frac{v^2}{2 D_z({\bf r})}\right],
\label{ksicloud}
\end{eqnarray}
where the correlation function $\xi_s$ is defined in the PPV space, where $R$ is the spatial separation
between points in the plane-of-sky and $v$ is the separation along the V-axis.
\begin{table*}[htb]
\begin{center}
\begin{tabular}{lccc} \hline\hline\\
& \multicolumn{1}{c}{ 1D: $P_s(k_v)$} & 2D: $P_s(K)$ & 3D: $P_s(K,k_v)$  \\[2mm]
& \multicolumn{1}{c}{ $ k_v D_z^{1/2}(S) \gg 1 $} & $ KS \gg 1 $
&  $ k_v^2 D_z(S) \gg (k S)^m $ 
\\[2mm] \hline \\
$ P_{\rho}: $ & $(r_0/S)^\gamma \left[k_v D_z^{1/2}(S)\right]^{2(\gamma-1)/m} $
& $ \left(r_0/S\right)^\gamma \left[K S\right]^{\gamma+m/2-3} $
&  $ (r_0/S)^\gamma \left[k_v D_z^{1/2}(S)\right]^{-2 (3-\gamma)/m} $ \\[2mm] 
\hline \\
$ P_{v}:$ & $ \left[k_v D_z^{1/2}(S)\right]^{-2/m}$
& $\left[K S\right]^{m/2-3}$
& $\left[k_v D_z^{1/2}(S)\right]^{-6/m} $ \\[3mm]
\hline
\end{tabular}
\end{center}
\caption{The short-wave asymptotical behavior
of power spectra  $P_s$ in PPV space. Results are presented for one dimensional spectrum of fluctuations along the velocity coordinate
$P_s(k_v)$, two-dimensional spectrum of fluctuations $P_s(K)$, three-dimensional anisotropic spectrum of PPV fluctuations
$P_s(K, k_v)$. We use the convention that capital letters denote 2D vectors in PP-plane.The adopted convention is that the variables related to spatial coordinates are denoted with capital letters. The component of $P_s$ arising from pure density fluctuations is  $P_\rho$, while the component affected both by the density and velocity fluctuations is $P_v$.  The power-law underlying statistics of density and velocity are assumed: 
$\gamma$ is the spectral index of the density correlation
function, $m$ is the spectral index of the velocity correlation function. The size of the turbulent cloud is $S$. From LP06.}
\label{tab:1Dspk_asymp}
\end{table*}

The correlation function of over-density given by  Eq.~(\ref{xifirst}) has a constant part that depends on the mean density
only and a part that changes with $r$. For instance, for the power-law
density spectrum the correlation functions of over-density take the form (see LP06
for the discussion of cases of $\gamma<0$ and $\gamma>0$):
\begin{equation}
\xi(r)= \langle \rho \rangle^2 
\left(1 + \left[ {r_0 \over r} \right]^\gamma\right), 
\label{Appeq:xi}
\end{equation}
where $r_0$ has the physical meaning of the scale at which fluctuations
are of the order of the mean density (see more in LP06). Substituting Eq.~(\ref{Appeq:xi})
in Eq.~(\ref{ksicloud}) it is easy to see that
 the PPV correlation
function $\xi_s$ can be presented as a sum of two
terms, one of which does depend on the fluctuations of density, the other does not.
Taking Fourier transform of $\xi_s$ one gets the PPV spectrum $P_s$, which is also
a sum of two terms $P_{\rho}$ and $P_{v}$, namely:
\begin{equation}
P_s=P_\rho+P_v~~,
\label{ps}
\end{equation}
where the assymptotics for $P_{\rho}$ and $P_{v}$ in one dimension (along the velocity axis), two dimensions (in the velocity slice) and three dimensions (the entire PPV space) are presented in Table~1. 

It is worth noting that while in the expressions for density and velocity correlations, i.e. Eq.~(\ref{Appeq:xi}) and (\ref{rm}), the spectral indexes of
the $\gamma$ and $m$ enter the same way. However, they enter the expression for $P_v$, which is the part of the spectrum affected both by density
and velocity, in very different ways. The origin of this difference can be seen in the expression for the correlation function of PPV intensity 
given by Eq.~(\ref{ksicloud}).  This is a mathematical consequence of the effect illustrated by Figure~\ref{eddies}. Depending on the values of $\gamma$ and $m$ either $P_\rho$ or $P_v$ dominates in the assymptotical regime. 

Expression (\ref{ksicloud}) and its generalizations
may be used directly to solve the inverse problem to find the properties of the underlying astrophysical
turbulence for an arbitrary spectrum.
 However, most attention so far was given to the astrophysically important case of
power-law turbulence. Note, that using spectra rather than the 
correlation function has advantages. For instance, the correlations along the V-axis of the PPV cube may be
dominated by large-scale gradients, while spectra provide correct result\footnote{Modified correlation functions
suggested in LP08 take care of the large-scale gradients and also provide correct results for the VCS studies.} (see
explanation in LP04 and LP06).  The results for 1, 2 and 3 dimensional spectra are presented in
Table~1. 

Whether $P_\rho$ or $P_v$ dominates depends on the statistical properties of density given by the spectral index $\gamma$. It was shown in LP07
that for $\gamma<0$ the $P_\rho$ contribution is always subdominant. For $\gamma>0$ Table~2 presents special
cases when the contribution of $P_\rho$ exceeds that of $P_v$. For instance, two first lines of the Table~2 define for a range of values $\gamma$ the
minimal velocity for which $P_\rho>P_v$.

Above we did not consider the galactic rotational velocities, which can be large. The justification for this can
be most easily understood if one deals with a spectral representation of the statistics. 
Taking Fourier transforms we deal with velocity
gradients, which are larger for turbulent motions than for large-scale sheer. For instance,
the latter
for the Galactic rotation is given by the Oort's constant, which is $14$~km~s$^{-1}$ kpc$^{-1}$.
In comparison, the shear due to typical Kolmogorov-type
turbulent motions in the Galaxy with the injection
of energy at 10 km~s$^{-1}$ at the scale of $L\sim 30$~pc is $\sim 300$~km~s$^{-1} (L/l)^{2/3}$ kpc$^{-1}$. Thus, in spite of the fact, that
regular large-scale galactic shear velocities may be much larger than the turbulent velocities, 
they can be neglected for our analysis (LP00 and a numerical study in Esquivel et al. 2003). 

In addition, the simplified discussion above ignored the self-absorption of the radiation. 
The intensities $I_{\bf X}(v)$, in general, are affected by both turbulence and
 absorption. To quantify these effects one can
consider the standard equation of radiative transfer
\begin{equation}
dI_{\nu}=-g_{\nu} I_{\nu} ds+j_{\nu}ds~~~,
\label{absor}
\end{equation}
where, for absorption and emissivity coefficients $\alpha$ and $\tilde{\epsilon}$, $g_{\nu}=\alpha({\bf x}) \rho({\bf x}) \phi_v({\bf x})$,
$j_{\nu}=\tilde{\epsilon} \rho({\bf x}) \phi_v({\bf x})$, ${\bf x}$ is a three
dimensional position vector $({\bf X}, z)$, $\rho({\bf x)}$ is the density and 
$\phi_v({\bf x})$ is the 
velocity distribution of the atoms. The Eq.~(\ref{absor}) is the starting point of a
detailed discussion in LP04, LP06, LP08. 

Other complications include the geometry of lines of sight and the dependences of the emissivities on
the density squared. Indeed, Eq.~(\ref{ksicloud}) assumes that the emission is collected along the parallel lines
of sight, which is an approximation valid for studying turbulence in a distant cloud, but not valid for studying turbulence
in gas at high galactic latitudes or in nearby regions of space. The corresponding generalization is provided in
Chepurnov \& Lazarian (2008b). There also the expressions in Table~1 are modified for the case of emissivities $\sim \rho^2$, as discussed in \S 8. There we also show that the practical data handling may provide better accuracy if one uses fitting of the PPV observational data with turbulence models, where the energy injection scale and the temperature of gas are used as additional fitting parameters. General expressions obtained in the aforementioned theoretical studies should be used for this purpose.   

\begin{table}[t]
\begin{tabular}{llc}
\hline
$ m \ge \mathrm{max}\left[\frac{2}{3},\frac{2}{3}(1-\gamma)\right] $
 & $ v^2 <  D_z(S) (r_0/S)^m $ \\
$ \frac{2}{3}(1-\gamma) < m < \frac{2}{3} $
& $ v^2 < D_z(S) (r_0/S)^{\frac{2/3 \gamma m}{m-2/3(1-\gamma)}} $ \\
$ m \le \mathrm{min}\left[\frac{2}{3},\frac{2}{3}(1-\gamma)\right] $
& $ r_0/S > 1 $ 
\end{tabular}
\caption{Conditions for the impact of
density inhomogeneities to the PPV statistics exceeds
the velocity contribution. Spectral index of density fluctuations $\gamma$ must be larger than 0, i.e.
the intensity of fluctuations increases with the decrease of scale. For $\gamma<0$ the velocity fluctuations
always dominate in creating small-scale ripples in the PPV space. From LP06.}
\label{table:density}
\end{table}
%
\section{Illustration of VCS and VCA Techniques for power-law velocities and densities}
Turbulence at its inertial range exhibits power-law spectra of velocity and density (see
Biskamp 2003 and ref. therein). Therefore, it is natural to study the relation between the PPV
statistics and power-law underlying statistics of velocity and density.

\subsection{Velocity Channel Analysis}
The interpretation of the channel maps is the domain of the VCA. 
Table~3 shows how the power spectrum of the intensity fluctuations depends on the
thickness of the velocity channel. Below we provide quantitative discussion of the VCA.
\begin{table*}
\begin{center}
\begin{tabular}{lcc}
\hline
Slice & Shallow 3-D density & Steep 3-D density\\
thickness & $P_{n} \propto k^{-3+\gamma}$, $\gamma>0$ &$P_{n} \propto k^{-3+\gamma}$, $\gamma<0$\\
\hline
2-D intensity spectrum for thin~~slice &
$\propto K^{-3+\gamma+m/2}$    & $\propto
K^{-3+m/2}$   \\
2-D intensity spectrum for thick~~slice & $\propto K^{-3+\gamma}$
& $\propto K^{-3-m/2}$  \\
2-D intensity spectrum for very thick~~slice & $\propto K^{-3+\gamma}$ & $\propto \
K^{-3+\gamma}$  \\
\hline
\end{tabular}
\end{center}
\caption{The VCA assymptotics. {\it Thin} means that the 
channel width $<$ velocity dispersion at the scale under
study;
{\it thick} means that the 
channel width $>$ velocity dispersion at the scale under
study;
{\it very thick} means that a
substantial part of the velocity profile is integrated over.
}
\end{table*}
It is easy to see that both for steep and shallow underlying density
the power law index
{\it steepens} with the increase of velocity slice
thickness. In the thickest velocity slices the velocity information
is averaged out and we get the
density spectral index $-3+\gamma$. The velocity fluctuations dominate in
thin slices, 
and the index $m$ that characterizes the velocity  fluctuation
can be obtained using thin velocity slices (see Table~1). As we mentioned earlier, the
notion of thin and thick slices depends on the turbulence scale under
study and the same slice can be thick for small scale turbulent fluctuations
and thin for large scale ones (see Figure~\ref{fig1}).

One may notice that the spectrum
of intensity in a thin slice gets shallower as the underlying
velocity get steeper. To understand this effect let us consider turbulence
in  an incompressible optically thin medium. The intensity of emission
in a velocity slice is proportional to the number of atoms per
velocity interval given by the thickness of the slice.
Thin slice means that the velocity dispersion at the scale of
study is larger than the thickness of a slice. The increase
of the velocity dispersion at a particular scales means that
less and less energy is being emitted within the velocity
interval that defines the slice (see Figure~(\ref{eddies})).  Mathematically
this effect results in the dependences in Table~3.

 If density variations are
also present they modify this result. However,
for small-scale asymptotics of thin slices this happens only when the density spectrum
is shallow (i.e. $\gamma>0$), i.e. dominated by fluctuations at small scales (see Eq.~(\ref{Appeq:xi})).

\subsection{Velocity Coordinate Spectrum}
The VCS is a brand new technique, which, unlike the VCA, was not motivated by the 
interpretation of the
existing observations. In the case of the VCS it was theoretical advances that induced the 
subsequent data analysis.

Unlike the standard spatial spectra, that are functions of angular wavenumber, the VCS is a function of
the wave number $k_v\sim 1/v$, which means that large $k_v$ correspond to small velocity differences,
while small $k_v$ correspond to large velocity differences. 
\begin{figure*}
\hbox{
  \includegraphics[height=.3\textheight]{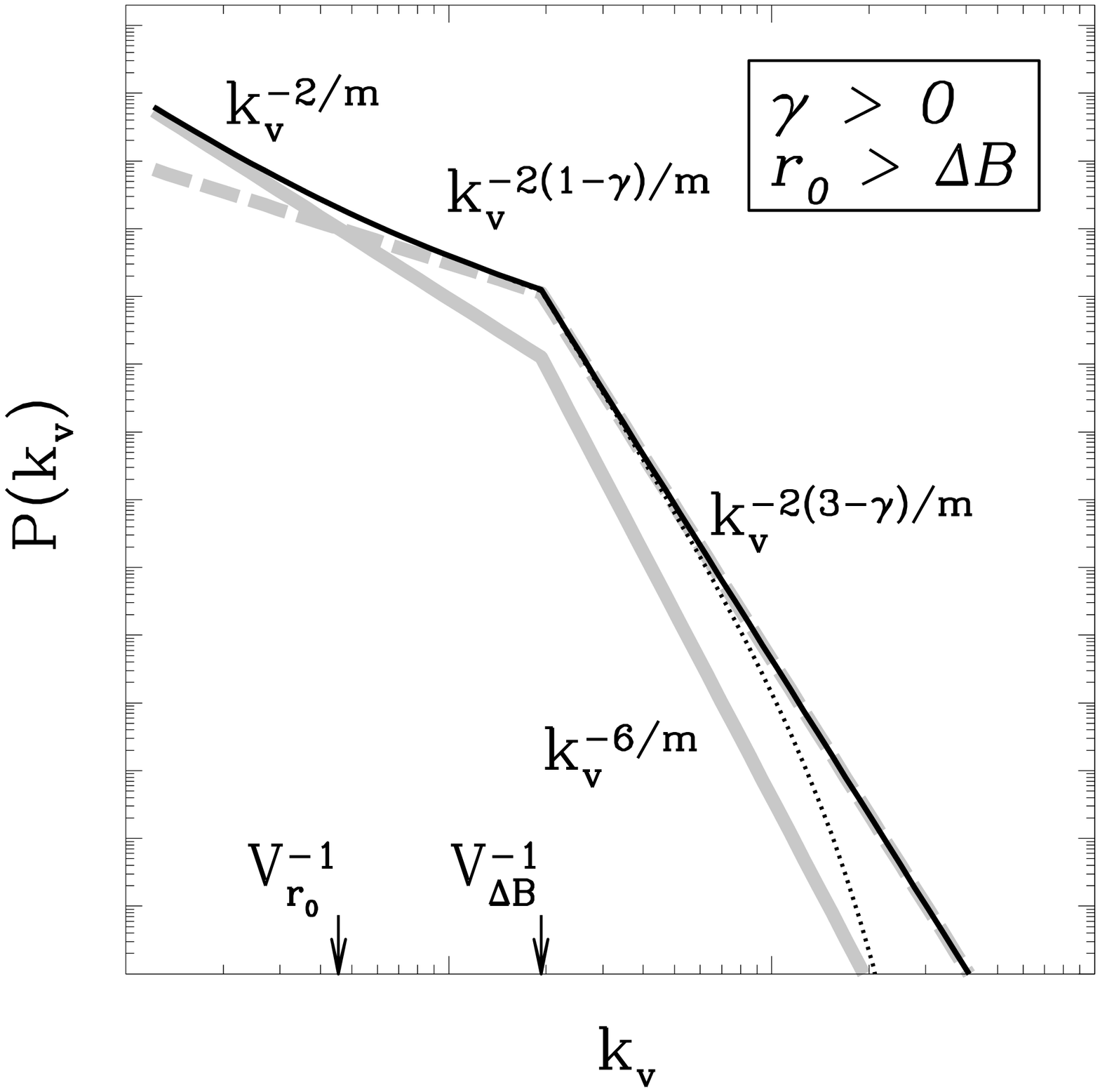}
\hfill
 \includegraphics[height=.3\textheight]{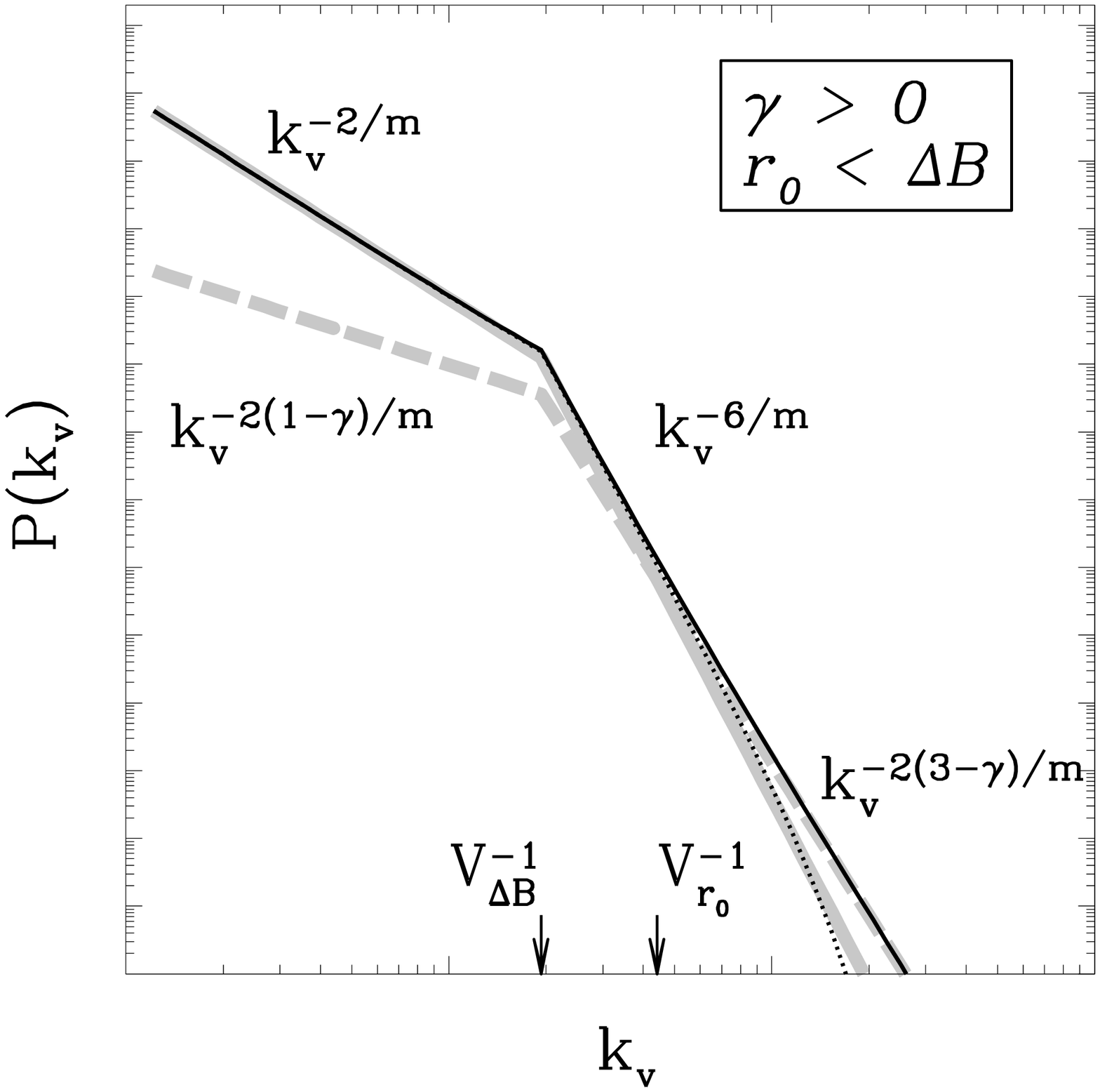}
\hfill
 \includegraphics[height=.3\textheight]{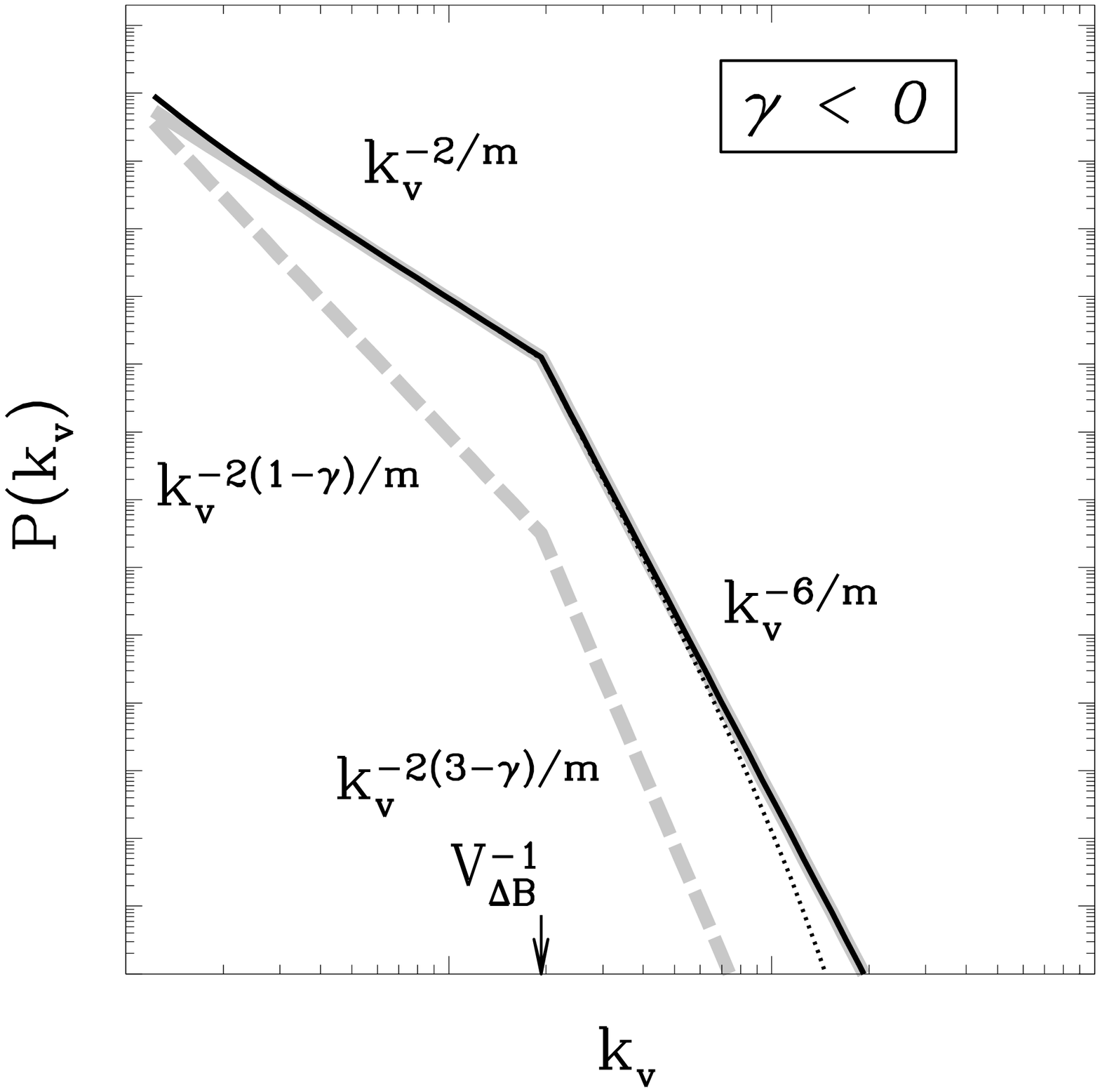}
}
  \caption{Particular cases of studying turbulence with the VCS technique.
In every panel light lines show contributions from the $\rho$-term
(density modified by velocity, dashed line) and $v$-term (pure velocity
effect, solid line) separately, while the dark solid line shows the combined
total VCS power spectrum. Thermal suppression of fluctuations
is shown by the dotted line. 
Labels below the dark solid lines mark the scaling of the subdominant
contributions.
For the {\it left} and {\it middle panels} the density
power spectrum is taken to be shallow, i.e., $\gamma > 0$.
The left panel corresponds to high amplitude of the density correlations,
$r_0 > \Delta B$, i.e. when density effects become dominant at relatively
long wavelengths for which the beam is narrow. In the middle panel,
the amplitude of density correlations is low $r_0 < \Delta B$ and they
dominate only the smallest scales which results in the intermediate steepening
of the VCS scaling. The {\it right panel} corresponds to the  steep density spectrum.
In this case the density contribution is always subdominant.
In this example the thermal scale is five times shorter than
the resolution scale $V_{\Delta B}$. From LP06.}
\label{case}
\end{figure*}

Assume that the maximal
resolution of a telescope corresponds to resolving the scale $\Delta B$
of a cloud at distance $L_{cloud}\gg S$.   At this scale the turbulent velocity is $
V_{\Delta B} \equiv \sqrt{D(S) (\Delta B/S)^m}$. It is not difficult to see that when
$
k_v^{-1} > V_{\Delta B} 
$
the beam is narrow, while for observations of smaller scales its width is important. 
The spectrum of fluctuations along the V-coordinate at the scale $k_v$ depends on whether the instrument  
resolves the correspondent spatial scale
$\left[k_v^2 D_z(S)\right]^{-1/m} S$. 
If this scale is resolved then $P_v(k_v) \propto k_v^{-2/m}$
and $P_{\rho}(k_v) \propto k_v^{-2(1-\gamma)/m}$. 
If the scale is not resolved then
$P_v(k_v) \propto k_v^{-6/m}$ and $P_\rho(k_v) \propto k_v^{-2(3-\gamma)/m}$. 
These results are presented in a compact form in Table~\ref{table:results}.
\begin{table}[h]
\begin{tabular}{lll} \hline\hline \\[-2mm]
Spectral~term & $\Delta B < S \left[k_v^2 D_z(S)\right]^{-\frac{1}{m}}$ &
$ \Delta B > S \left[k_v^2 D_z(S)\right]^{-\frac{1}{m}}$ \\[2mm]
\hline \\[-2mm]
$ P_\rho(k_v) $ & $ \propto\left(k_v D_z^{1/2}(S)\right)^{-2(1-\gamma)/m} $& 
$ \propto\left(k_v D_z^{1/2}(S)\right)^{-2(3-\gamma)/m} $ \\[2mm]
\hline \\[-2mm]
$ P_v(k_v) $ & $ \propto\left(k_v D_z^{1/2}(S)\right)^{-2/m} $  & 
$ \propto\left(k_v D_z^{1/2}(S)\right)^{-6/m} $ \\[2mm]
\end{tabular}
\caption{Scalings of VCS for shallow and steep densities for measurements
taken with the telescope with a finite beam size. From LP06. }
\label{table:results}
\end{table}

The transition from the low to the high resolution regimes happens as
the velocity scale under study gets comparable to the turbulent velocity
at the minimal spatially resolved scale. As the change of slope is a
velocity-induced effect, it is not surprising that the difference in
spectral indexes in the low and high resolution limit is $4/m$ for both the 
$P_v$ and $P_\rho$ terms, i.e it does not depend on the density\footnote{In
the situation where the available telescope resolution is not sufficient,
i.e. in the case of extragalactic turbulence research, the high spatial
resolution VCS can be obtained via studies of the absorption lines from
point sources.}.
This allows for separation of the velocity and density contributions.
For instance, Figure~\ref{case}
illustrates that in the case of shallow density both the density and velocity 
spectra can be obtained. 
Potentially, procedures for extracting 
information on 3D turbulent
 density can be developed for the steep density case
as well. However, this requires careful accounting for errors 
as the contribution from density is subdominant in this case\footnote{Needless to say, when the turbulent object is resolved, the 
easiest way to obtain the density
spectral index is to study the integrated intensity maps, provided that the
absorption is negligible (see criteria for this in LP04).}.

\subsection{Effects of Self-Absorption}
The issues of absorption were worrisome for researchers from the
very start (see Munch 1958). Unfortunately, erroneous
statements about the effects of absorption on the observed turbulence
statistics are widely spread (see discussion in LP04).

Using transitions that are less affected by absorption, e.g. HI,
may allow us to avoid the problem. However, it is regretful not
to use the wealth of spectroscopic data only because absorption
may be present. A study of absorption effects
in is presented in LP04 and LP06. For the VCA it was found that for sufficiently thin\footnote{The thermal 
broadening limits to what extent the slice can be thin. This means that in some cases
that the actual turbulent velocity spectrum may not be recoverable.} slices
the scalings obtained in the absence of absorption still hold
provided that the absorption on the scales under study is negligible.

When dealing with self-absorption, one should start with Eq.~(\ref{absor}) (see LP04).
The criterion for the absorption to be important is 
$\alpha^2 \langle (\rho_s({\bf X}, v_1)-\rho_s({\bf X},v_2))^2\rangle\sim 1$, which for $\gamma<0$
 results in the critical size of the slice thickness $V_c$ given by (LP06)
\begin{eqnarray}
V_{c}/D_z(S)^{1/2}
 &\approx& \left(\alpha \bar \rho_s\right)  ^{\frac{2m}{m-2}}, 
~~~~~~ m > 2/3 \label{eq:abs_width1}\nonumber \\
V_{c}/D_z(S)^{1/2}
 &\approx& \left(\alpha \bar \rho_s\right)^{-1}~,
~~~~~~ m < 2/3,
\label{eq:abs_width2}
\end{eqnarray}
where $\bar \rho_s$ is the mean PPV density.
The absorption is dominant for the slices thicker than $V_c$. The difference with the case of
$\gamma>0$ is that, in the latter case, one should also consider the possibility that the density 
contribution can be important (see Table~2).
The criterion above coincides with one for the VCS, if we identify the critical
$k_v$ with $1/V_c$. If the resolution of the telescope is low, another limitation applies.
The resolved scale should be less than the critical spatial scale that arises from
the condition $\alpha^2 \langle (\rho_s({\bf X_1}, v)-\rho_s({\bf X_2},v))^2\rangle\sim 1$ which
for $\gamma<1$ results in $R_{c}/S \approx 
\left(\alpha \bar \rho_s\right)  ^{\frac{2}{m-2}}$ (LV06). If only  
scales larger than $R_c$ are resolved, the
information on turbulence is lost.
     
If integrated intensity of spectral lines is studied in the presence of
absorption non-trivial effects emerge. Indeed, for optically thin
medium the spectral line integration results in PPV intensity fluctuations that reflect
the density statistics. LP04 showed that this may not
be any more true for lines affected by absorption.
When velocity is dominant a very interesting regime for which
intensity fluctuations show universal behavior, i.e. the
power spectrum $P(K)\sim K^{-3}$  emerges.
When density is dominant (see Table~2),
the spectral index of intensity fluctuations in those two situations is the same
as in the case of an optically thin cloud integrated through its volume. This  means
that for $\gamma>0$, i.e. for steep spectrum of density, in the range of parameter space defined by Table~2 
the measurements of intensity fluctuations of the integrated spectral
lines reflect the {\it actual} underlying density
spectrum in spite of the absorption effects.

\section{Studies of turbulence with absorption lines}

\begin{figure}
  \includegraphics[width=.45\textwidth]{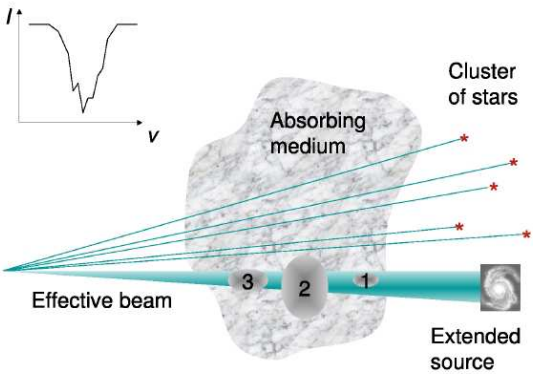}
  \caption{Illustration of VCS absorption studies of turbulence. {\it Left Panel}: Schematic of measuring turbulence
  with absorption lines from point sources, e.g. stars, and extended source, e.g. a galaxy. 
From Chepurnov \& Lazarian 2008.}
\label{abs10}
\end{figure}

The analysis of weak absorption lines is analogous to the analysis of the weak emission lines. For instance, the weak absorption
data from extended sources, e.g. of atomic gas towards Cassiopea A  and Cygnus A in Deshpande et al. (2000), can be interpreted with the traditional VCA technique. VCS can handle both the weak absorption data from both point sources and extended sources (see Figure \ref{abs10}).

New effects arise when strong absorption lines, which are in a saturated regime, are studied. This problem was addressed in Lazarian \& Pogosyan (2008, henceforth LP08), who proposed to analyze optical depth fluctuations. These correspond to the logarithms of the measured intensities. The analysis was made for the VCS technique, as this is the technique that requires sampling of a turbulent volume just over a few lines of sight. Indeed, numerical simulations in Chepurnov \& Lazarian (2008) proved that the sampling along from 5 to 10 directions to absorbing sources, e.g. stars, is enough to recover properly the underlying spectrum of turbulent velocity (see Figure \ref{abs1}).
 
At what optical depth the is the recovery of the turbulence spectra feasible? It is well known that for optical depth $\tau$ larger than 
$10^5$ the wings are totally dominated by Lorentz factor (see Spitzer 1978). For the range of optical depth less than $10^3$, the line width is determined by Doppler shifts rather than the atomic constants, which simplifies the study.
 While formally the entire line profile 
provides information about the turbulence, in reality, the flat saturated part of the profile will contain only noise and will not be useful for any statistical study. Thus, it is the wings of saturated absorption lines that can be used for turbulence studies.

If however, the absorption lines are studied using an extended emission source, then the VCA analysis is also possible with the
logarithms of the intensities in the velocity channels. For both the VCA and VCS techniques, the effect of saturation of the absorption line results in limiting the range of $k_v$ at which the information on turbulence is available. If one approximates the effect of
saturation with the help of a Gaussian mask of width $\Delta$, centered in the middle of the wing, $\Delta$ measures the fraction
of the line that is available for studies of turbulence. Our studies (see Figure \label{satur}) in LP08 show that the recovery of
the turbulent spectrum with the VCS is possible for $k_v>3\Delta^{-1}$, where all the quantities are normalized over the total turbulent velocity dispersion. In terms of the VCA this translates into the requirement that
the thickness of the velocity channel over which it is feasible to analyze the spectrum of the {\it logarithm} of the intensity fluctuations
is $<1/3 \Delta$. Thermal effects provide additional limitations for the range of scales available for observations, i.e. for finite 
$\beta'=\beta/D_z(S)$ (see Eq. \ref{phi}), where $D_z(S)$ is the velocity dispersion of the line the range of $k_v$ for studies
of asymptotic power-law solutions is limited to $3\Delta^{-1}<k_v<1/(3\beta)$. For $k_v$ beyond this range, the recovery of the turbulence spectrum is still possible, but fitting of the integral expressions, rather than the use of the asymptotical solutions is necessary (see \S 8).  

\begin{figure}
  \includegraphics[width=.45\textwidth]{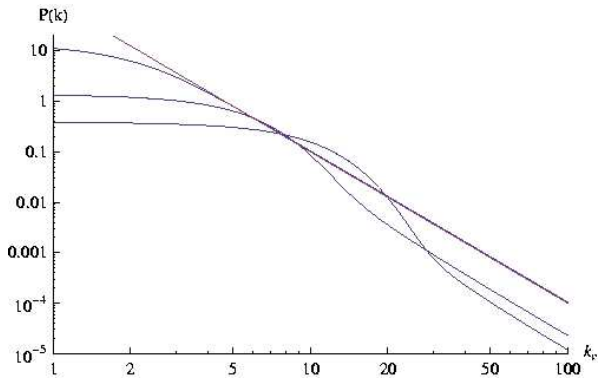}
  \caption{Power spectrum of the optical depth fluctuations from line wings.
All parameters, $k_v$, are dimensionless,
in the units of the variance of the turbulent velocity at
the scale of the cloud.  
Only the effect of the turbulent motions and not spatial inhomogeneity of the
absorbers is taken into account. The underlying scaling of the turbulent
velocities is Kolmogorov, $m=2/3$.
The left panel illustrates the power aliasing due to finite width of the window.
The power spectrum is plotted, from top to bottom, for $\Delta=1,0.2.0.1$, i.e
the widths of the wing ranges from the complete line to one-tenth of the
line width. Effects of thermal broadening that distort the spectrum at large $k_v$ are not shown.  From LP08.}
\label{satur}
\end{figure}

When several absorption lines from different species are available along the same line of
sight, one can improve the recovery of the turbulence spectrum by combining
them together. We believe that piecewise analyses of the wings belonging
to different absorption lines is advantageous. Optical and UV absorption lines
are the primary targets for such an analysis.
 Formally, for lines with weak absorption,
i.e. $\tau_0<1$, there is no need for other measurements.
However, in the presence of inevitable noise, the situation may be far from
trivial. Naturally, noise of a constant
level, e.g. instrumental noise, will affect more weak absorption lines. 
The strong absorption lines, in terms of VCS, sample turbulence only for
sufficiently large $k_v$. This limits the range of turbulent scales that can
be sampled with the technique. However, the contrast that is obtained with the
strong absorption lines is higher, which provides an opportunity of increasing
signal to noise ratio for the range of $k_v$ that is sampled by the absorption
lines.

\section{Comparison of VCA and VCS}
 
Traditionally the techniques to study velocity
turbulence, e.g. velocity centroids or VCA, require observations to spatially resolve the scale
of the turbulence under study\footnote{As it was discussed in LP00, the VCA
can be applied directly to the raw interferometric data, rather than to
images that require good coverage of all spatial frequencies. However, even
with interferometers, the application of the VCA to extragalactic objects is
restricted.}. This constrains the variety of astrophysical
objects where turbulence can be studied.
In this way, the VCS,
is a unique tool that allows studies of astrophysical turbulence even when
the instrument does not resolve spatially the turbulent fluctuations.

\begin{figure}
 \begin{center}
 \includegraphics[width=.5\textwidth]{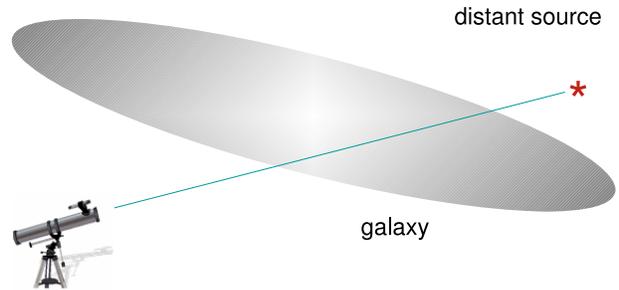}
 \includegraphics[width=.5\textwidth]{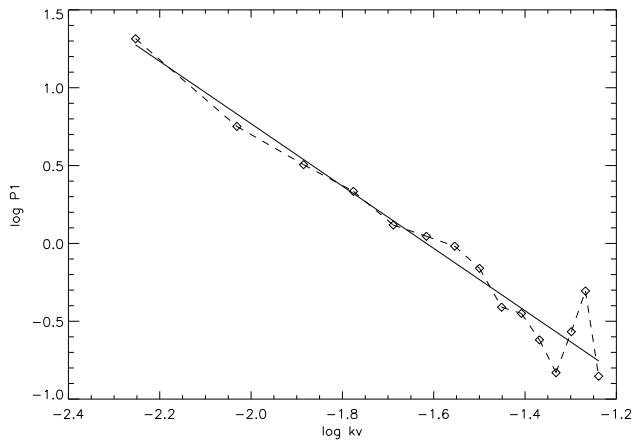}
 \end{center}
  \caption{{\it Top Panel}: Studies of turbulence using a single absorption spectral line. {\it Bottom Panel}: Velocity Coordinate Spectrum obtained
  using sampling of a turbulent volume along 10 lines of sight.  The solid line corresponds to the theoretical expectations. From Chepurnov \& Lazarian 2008.}
\label{shear}
\end{figure}

Can the VCS technique recover the turbulence spectrum while dealing with spatially unresolved astrophysical objects?
From a pure theoretical standpoint, this should be feasible. Indeed, if we deal with fluctuations at very small scales we
can identify different parts of the spectral line with different statistical realizations of the small-scale stochastic process.
This enables us to perform the averaging using those parts. From a practical point of view, thermal broadening limits the
range of the scales that can be resolved spectroscopically. Thus, reliable studies of turbulence may require spatial averaging. 
Chepurnov \& Lazarian (2008) established with numerical simulations that 5-10 measurements of the spectrum can
be sufficient for performing an adequate spatial averaging. This means that for a turbulent volume with marginal 
spatial resolution but good spectral resolution, studies of turbulence with the VCS are possible using emission lines.
In addition, if the turbulent volume rotates, the resulting Doppler broadened lines with width much larger than the turbulent
width can be chopped into spectral pieces that can be used to perform statistical averaging. This makes feasible the recovery
of the spectrum with a single spectral measurement (see Figure \ref{shear}).  

Our study of the effect of finite temperatures for the technique reveals that,
unlike the VCA, the temperature broadening does not prevent the turbulence
spectrum from being recovered from observations. Indeed, in VCA, gas
temperature acts in the same way as the width of a channel. Within the VCS
the term with temperature gets factorized and it influences the amplitude
of fluctuations (LP06). One can correct for this term\footnote{To do this, one may
attempt to fit for the temperature that would remove the exponential
fall off in the spectrum of fluctuations along the velocity coordinate
(Chepurnov \& Lazarian 2006a).}, which also allows for a new
way of estimating the interstellar gas temperature.

Another advantage of the VCS compared to the VCA is that it reveals the
spectrum of turbulence directly, while within the VCA the slope of the spectrum
should be inferred from varying the thickness of the channel. As the thermal
line width acts in a similar way as the channel thickness, additional care
(see LP04) should be exercised not to confuse the channel that is still
thick due to thermal velocity broadening with the channel that shows the 
thin slice asymptotics. A simultaneous
use of the VCA and the VCS makes the turbulence
spectrum identification more reliable.

The introduction of absorption in VCS and VCA brings about different results.
Within the analysis of velocity slices spectra (VCA) the absorption results
in new scalings for slices for which absorption is important. 
The turbulence spectral indexes 
can be recovered for the VCA within 
sufficiently thin slices, provided that the thickness of the slices
exceeds the thermal line width. For the VCS  at large $k_v$ for which absorption 
becomes important the spectra get exponentially damped. 

Both VCA and VCS are applicable to studies of not only 
emission, but also absorption lines.  However, the necessity of
using extended emission sources limits the extent of possible VCA studies of
turbulence. This is not an issue for the VCS, for which absorption 
lines from {\it point sources} can be
used (see Figure~\ref{abs10}). Interestingly enough, in this case the asymptotics for the high
resolution limit for the VCS technique should be used irrespectively of the actual beam size of the
instrument.

 \section{Numerical testing}
 
  \begin{figure*}
\begin{center}
\includegraphics[width=.45\textwidth]{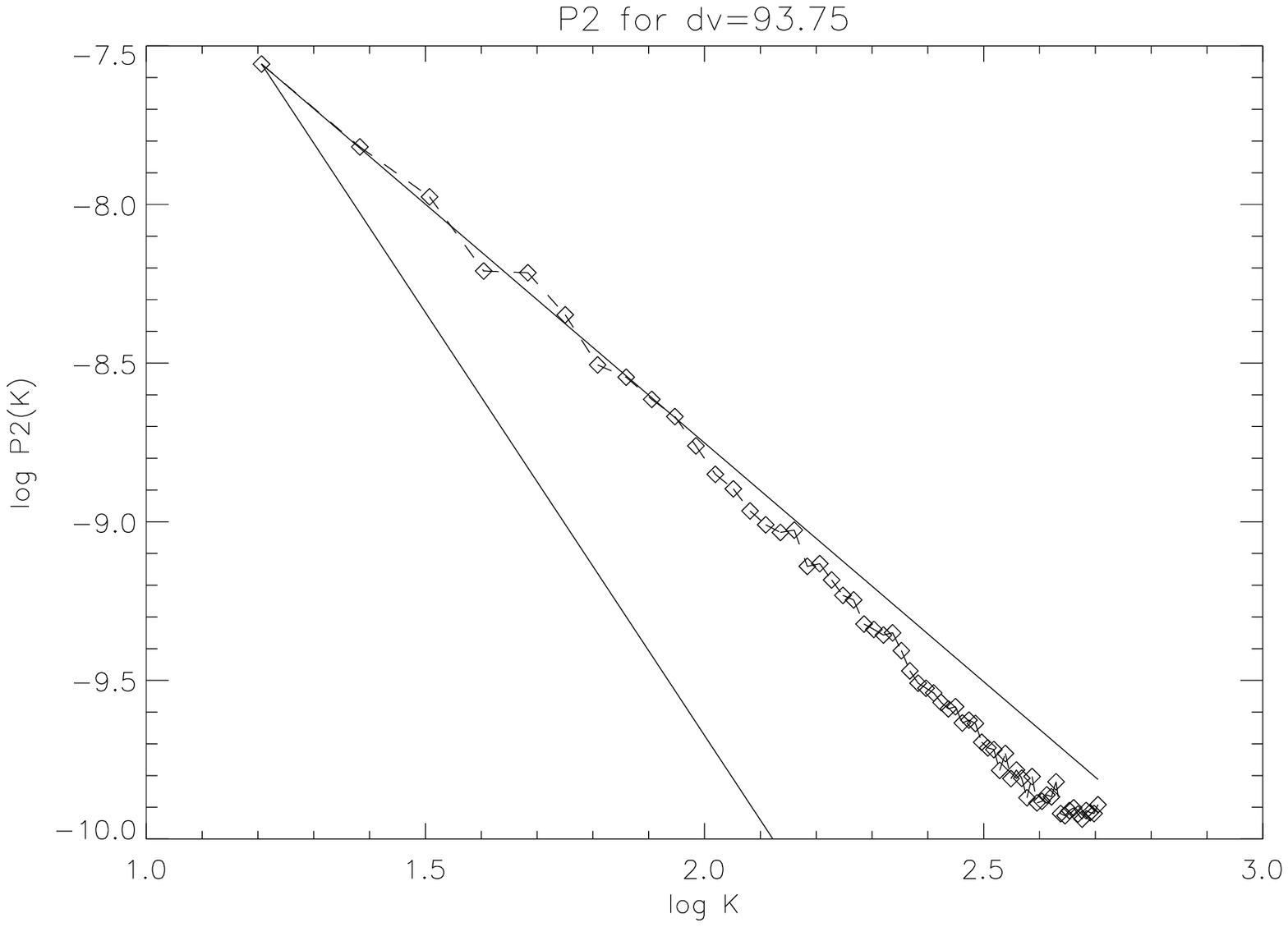}
 \includegraphics[width=.45\textwidth]{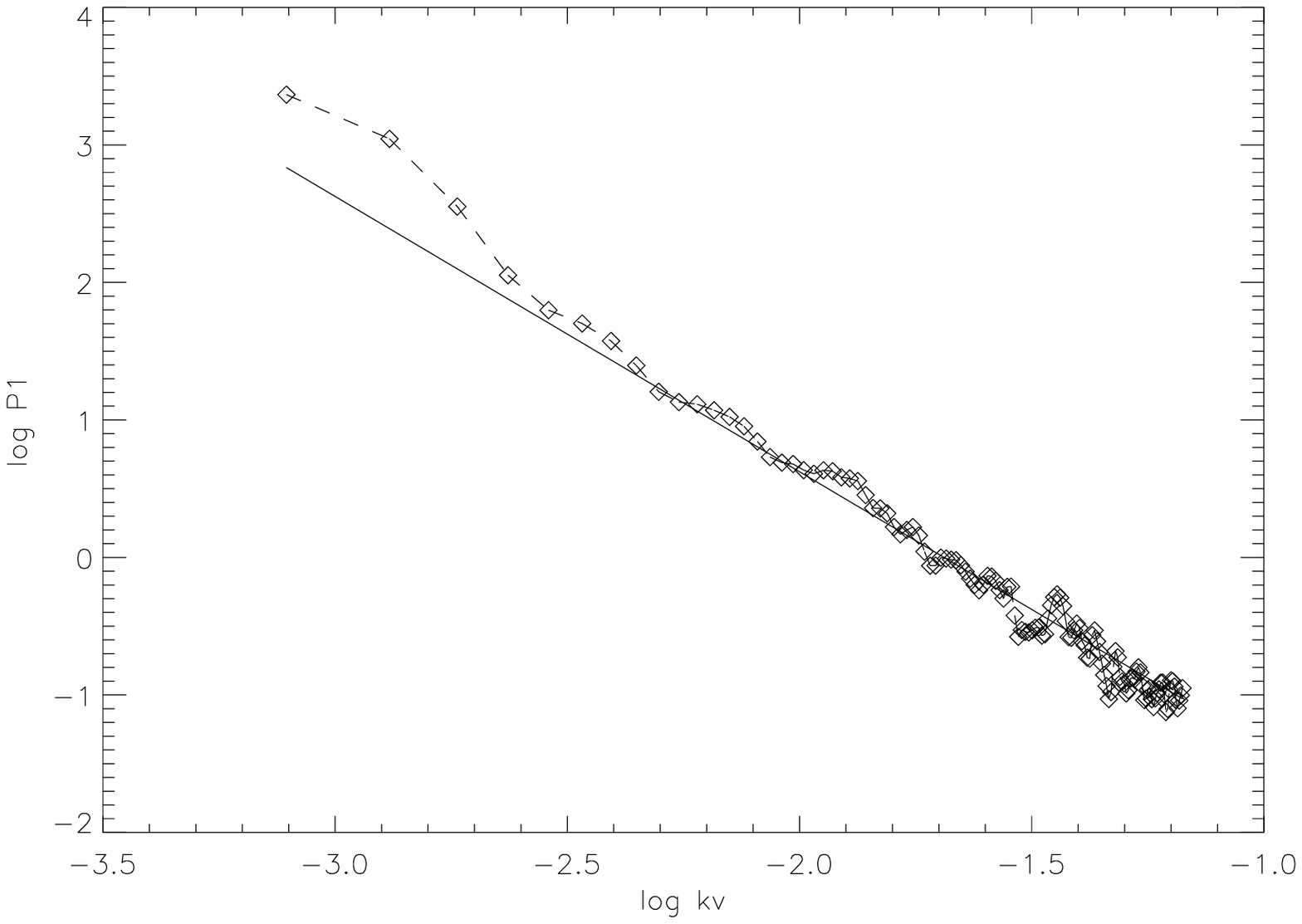}
\end{center}
\caption{Testing
of the predictions for the VCA and VCS techniques with synthetic observations. The underlying spectral index of velocity fluctuations is $-4$.
. {\it Left panel}: VCA spatial spectrum for different velocity slice thickness. Thin slice: shallower solid line shows the expected velocity-dominated spectrum, Thick slice: steeper solid line shows density-dominated spectrum. {\it Right panel}: Spectrum measured from the synthetic data with the VCS techniques versus the theoretical expectations.  
\label{VCA_VCS_4_00}}
\end{figure*}

VCA predictions were tested in Lazarian et al. (2001), Esquivel et al. (2003), Padoan et al. (2006)
and in Chepurnov \& Lazarian ( 2008) using synthetic maps obtained with 
synthetic power-law data as well as with numerical compressible MHD simulations.
Simulated data cubes allowed both density and velocity statistics
to be measured directly. Then these data cubes were
used to produce synthetic spectra which were analyzed using the
VCA. As the result, the velocity
and density statistics were successfully recovered.
 
The most extensive high resolution testing of VCA and the first numerical testing of VCS was performed in Chepurnov \& Lazarian (2008).
The results for the velocity spectrum corresponding to shocks, i.e. for the $E(k)\sim k^{-2}$, are shown in Figure~\ref{VCA_VCS_4_00}. The spectra for  
more shallow velocities, e.g. the Kolmogorov spectrum\footnote{Note, that in terms of three dimensional spectrum, Kolmogorov spectrum corresponds to $k^{-11/3}$. This difference arises 
from the averaging over directions in $k$-space.} $E(k)\sim k^{-5/3}$  show more noise, which increases at small scales. This noise originates at velocities corresponding to the 
velocities of the adjacent numerical points and
is caused by the discrete nature of the data set involved in the simulations. Therefore, this noise is not expected for the real-world smooth 
distribution of turbulent fluid. This noise is not observed while the actual astrophysical data is handled either. However, this is an important effect for 
{\it numerical} testing\footnote{Low resolution testing of
 VCA in Miville-Deschenes et al. (2003) had inadequate resolution and therefore
 brought erroneous results.}.

\begin{figure}
 \includegraphics[width=.45\textwidth]{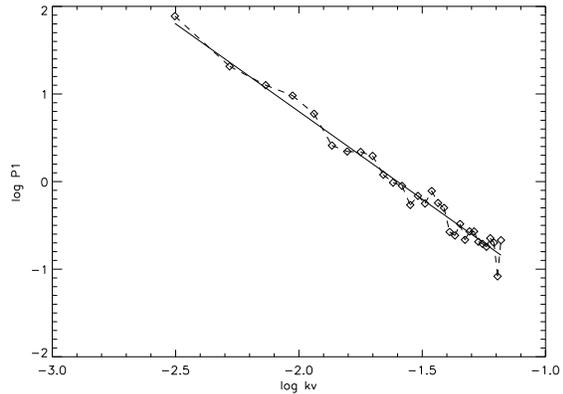}
  \caption{Spectrum of turbulence recovered from synthetic observations when the signal is sampled along
10 absorption lines. The solid line corresponds to the underlying spectrum and points correspond to the
recovered spectrum.
From Chepurnov \& Lazarian 2008.}
\label{abs1}
\end{figure}

Apart from testing of the VCA and VCS techniques, Chepurnov \& Lazarian (2008) tested the effect of
limited data samples on the noise in the spectra obtained. In particular, applying the VCS technique to 
synthetic observations of absorption lines, Chepurnov \& Lazarian (2008) showed that having just several
spectral absorption measurements is sufficient for
recovering the underlying turbulent velocity spectrum (see Figure~\ref{abs1}).

\section{Extending VCA and VCS}

\subsection{Studying non-power-law turbulence}

Historically, the VCA technique was initiated to explain the power-law dependences of emission in velocity channels reported by observers. These power-law dependencies are only approximate, as it follows from the theory (see LP00). Indeed, a velocity slice may contain the images of both small eddies, for which the slice is thick, and large eddies, for which the slice is thin (see Figure \ref{fig1}). Similarly, thermal broadening distorts the VCS output even for an underlying power-law turbulence. Needless to say, studies of the interstellar turbulence spectrum with injection and dissipation of energy at various scales are most exciting. This underlying spectrum is not a power-law and therefore the VCA and VCS data handling are not going to deliver power-law dependences either.\footnote{In general, studying turbulence dissipation is an exciting avenue for getting insight into the intimate details of the turbulence cascade. For instance, it has been shown numerically that MHD turbulence creates separate cascades of Alfvenic, fast and slow modes (Cho \& Lazarian 2002, 2003), with these modes having dissipation scales. This should result in the velocity spectrum deviating from the power law at small scales.}

While the analytical dependences that we presented in the Tables are very useful for understanding the nature the fluctuations observed and the initial analysis of the data, we believe that real progress in quantitative studies of turbulence can be achieved by using the integral expressions obtained in LP00, LP04, LP06, LP08 to fit the data with the turbulence models. The first work in this direction has been done with the VCS (see \S 9.2). We expect similar work to be done with the VCA.

\subsection{Studies of nearby gas}

The notion of "nearby gas" in terms of the turbulence studies is related to situations when the lines of sight coming from the telescope cannot be considered parallel. For instance, studies of gas in the Local Bubble may require taking into account the
convergence of the lines of sight.
Naturally, if the sampling happens along an infinitely thin line, there is no difference for the nearby gas and the distant gas in terms of the VCS technique. This is the case, for instance, for turbulence studies using absorption from a star. The spectral index for $P_v$ is given by Table~4, i.e. $P_v\sim k_v^{-2/m}$. 

If, on the contrary, the resolution is low, the difference between the distant object and nearby gas result in the 
a very different solutions for $P_v$, in particular, in the asymptotic of $P_v\sim k_v^{-4/m}$ (Chepurnov \& Lazarian 2008). Note, that this asymptotical spectrum is not as steep as the spectrum arising in the case of low resolution and parallel lines of sight (cf. Table~4). This simplifies recovering of the velocity spectrum from observations in the presence of noise.

\subsection{Studies of emission lines with emissivity $\sim \rho^2$} 

The quantities we deal with in spectral line observations are the velocity and the gas emissivity. Lazarian \& Pogosyan (see LP00, LP04, LP06, LP08) treated the emissivities proportional to the density to the first power. Therefore, in terms of scalings,
the emissivities and densities were interchangeable. This is not true, however, when the emissivities are proportional to $\rho^2$, as is the case of the recombination lines in plasma. The latter regime modifies the analysis. In particular, for the shallow spectrum of density, Chepurnov \& Lazarian (2008) showed that the spectral index of the correlation function of emissivity $\gamma_{\eps}=2\gamma$, where $\gamma$ is the index of the correlation function of density, which we discussed all the way above in the paper.  

The first consequence of this is that if the density is shallow the emissivity is shallow. The second consequence is that one can use the asymptotics the second column of Table~4, but substituting there $\gamma_{\eps}=2\gamma$ instead of $\gamma$. The same relation, i.e. $\gamma_{\eps}=2\gamma$ is applicable for shallow densities when the VCS is considered.

For steep density spectra the spectral index of emissivity $\gamma_{\eps}$ should be instead of $\gamma$ when the emissivities are proportional to 
$\rho^2$. This is true for both the VCA and VCS techniques, which once again stresses their close connection.

\section{Observational studies}

\subsection{Applying VCA}

\begin{table*}
\begin{center}
\begin{tabular}{|c|c|c|c|c|c|c|c|c|c|} \hline
N    & data          & Object                 &  $P^{thin}_{PPV} $     & $P_{PPV}^{thick}$                 & depth            & $E_v$               & $E_{\rho}$               & Ref. obs.             & Ref. theor.\\ \hline \hline
1     & HI             & Anticenter$^g$   & $K^{-2.7}$                  &         N/A                                  & Thin                & $k^{-1.7}$        & N/A                        & [1]                        & [2]             \\ \hline
2     &HI              &$\rightarrow$CygA &$K^{-(2.7)}$                &        $K^{-(2.8)} $                       & Thin               & N/A                    & $k^{-(0.8)}$             & [3]                        & [3]              \\ \hline
3     & HI             & SMC$^e$           & $K^{-2.7}$                  &        $K^{-3.4}$                        & Thin                 &  $k^{-1.7}$       & $k^{-1.4}$             & [4]                         & [4]             \\ \hline
4     & HI             & Center$^g$        & $K^{-3}$                     &        $K^{-3}$                            & Thick               & N/A                  & N/A                        & [5]                          & [6]             \\ \hline
5    & HI             & B. Mag.$^g$       &$K^{-2.6}$                  &        $K^{3.4}$                          & Thin                  &  $k^{-1.8}$        & $k^{-1.2}$            & [7]                          & [7]               \\ \hline
6     & HI             & Arm$^g$            & $K^{-3}$                     &        $K^{-3}$                            & Thick                & N/A                  & N/A                        & [8]                          & [9]             \\ \hline
7     & HI             & DDO 210$^e$    &$K^{-3}$                     &         $K^{-3}$                           & Thick                & N/A                   & N/A                        & [10]                          & [10] [9]          \\ \hline 
8    & $^{12}$CO  & L1512                &N/A                            &        $K^{-2.8}$                         & Thick                 &  N/A                  & $k^{-0.8}$             & [11]                         & [5]            \\ \hline
9    & $^{13}$CO & L1512                &N/A                             &        $K^{-2.8}$                          & Thick                 &  N/A                   & $k^{-0.8}$            & [11]                        & [10]            \\ \hline
10   & $^{13}$CO & Perseus              &$K^{-(2.7)}$               &        $K^{-3}$                              & Thick                 & $k^{-(1.7)}$       & N/A                      & [12]                         & [12]            \\ \hline
11 & $^{13}$CO  & Perseus            & $K^{-2.6}$                  &        $K^{-3}$                              & Thick                  & $k^{-1.8}$         & N/A                     & [13]                         & [13]            \\ \hline
12 & C$^{18}$O    & L1551                & $K^{-2.7}$                  &        $K^{-2.8}$                           & Thin                  & $k^{-1.7}$           & $k^{-0.8}$            & [14]                         & [14]           \\ \hline                                                     
\end{tabular}
\end{center}
\caption{{\small Selected VCA results. Superscript ``$^g$" denotes galactic objects, ``$^e$" -- extragalactic. $P^{thin}_{PPV}$ and $P_{PPV}^{thick}$ are the power law spectrum in thin and 
thick PPV slices, respectively. ``Ref. obs." and ``Ref. theor" correspond to papers where the measurement were done and interpreted using VCA, respectively. Indexes in round brackets
correspond to substantial observational errors correspond to consistency only. $\rightarrow$CygA is used to denote material towards Cygnus A.  [1] is Green (1993), [2] is Lazarian \& Pogosyan 
(2006), [3] is Deshpande et al. (2000), [4] is Stanimirovic \& Lazarian (2001), [5] is Dickey et al. (2001), [6] is Lazarian \& Pogosyan (2004), [7] is Muller et al. (2004), [8] is Khalil et al. (2006), [9] 
is Lazarian 2006, [10] is Begum et al. (2006), [11] is Stutzki et al. (1998), [12] is Sun et al. (2006), [13] is Padoan et al. (2006), [14] is Swift (2006).}}   
\label{table}
\end{table*}
Table~5 illustrates the results obtained with VCA by
different groups analyzing observations. It displays the variety
of objects to which VCA has been applied. The studies of spectra in
channel maps predated VCA (see lines 1, 8, 9 in Table~\ref{table} as
well as Crovisier \& Dickey 1983, Stanimirovic et al. 1999), but
researchers were choosing $\Delta v$ arbitrarily, making any sensible
comparisons impossible. For these cases, we performed the VCA analysis
using the published data. The results presented are based on the
detailed analytical treatment of different cases in Lazarian \&
Pogosyan (2000, 2004). For instance, Lazarian \& Pogosyan (2004)
predict that absorption can induce a universal\footnote{This assumes that the
velocity spectrum $E_v\sim k^{-\beta}$ has 
$\beta>1$. However, this is true for all turbulence spectra that we know.} spectrum $\sim K^{-3}$,
where $K$ is the observational analog, in the $2D$ plane of the sky, of
the wavenumber $k$. If this spectrum dominates in both thin and thick
slices, the only conclusion that can be made is that the density
spectrum $E_{\rho}\sim k^{-\alpha}$ corresponds to $\alpha>1$ (see
lines 3, 6, 7), while the details of the spectrum are not
available. When the spectrum of intensities in a thick slice is
different from $K^{-3}$ in the presence of absorption, the information
about the underlying densities is available (see lines 8, 9). It is
encouraging that the observed spectral indexes correspond to those in
simulations (see Beresnyak, Lazarian \& Cho 2006, Kowal, Lazarian \&
Beresnyak 2007), which show a tendency of having the spectrum of density
getting flatter as the spectrum of velocity gets steeper. A stronger
statement about the good quantitative correspondence between the VCA
analysis of Perseus data and their numerical simulations is made in
Padoan et al. (2006) and also in Kritsuk et al. (2007).

One of the first applications was in
Stanimirovic \& Lazarian (2001), where the technique was applied to the Small Magellanic Cloud (SMC) data.
The analysis revealed spectra of 3D velocity fluctuations roughly consistent with
the Kolmogorov scaling (a bit more shallow).  Esquivel et al. (2003)
used simulations of MHD turbulent flows to show that, in spite
of the presence of anisotropy caused by magnetic field, the
expected scaling of fluctuations is indeed Kolmogorov. Studies by
Cho \& Lazarian (2002, 2003) revealed that the Kolmogorov-type
scaling is also expected in the compressible MHD flows. This also
supports the conclusion in LP00 that the data in Green (1993)
is consistent with MHD turbulence scaling.

Studies of turbulence are more complicated for the inner parts of the Galaxy,
where (a) two distinct regions at different distances from the observer
contribute to the emissivity for a given velocity and (b) effects of
the absorption are important. However, the analysis in Dickey et al. (2001)
showed that some progress may be made even in those unfavorable
circumstances. Dickey et al. (2001) found the steepening
 of the spectral index with the increase of the velocity slice thickness.
They also observed the spectral index for strongly absorbing direction
approached $-3$ in accordance with the conclusions in LP04.
Note,  that  the effects of optical depths may explain
some other case when the spectral index stayed the same, e.g. -3, while the thickness of the slice
was varying (see Khalil et al. 2006). Incidently, this situation
 can be confused with the situation when 
the fluctuations arise from density only (see Begum et al. 2006).

21-cm absorption provides another way of probing turbulence on small
scales. The absorption depends on the density to temperature ratio
$\rho/T$, rather than to $\rho$ as in the case of emission\footnote{
In the case of an isobaric medium the product of
density and temperature are constant and the problem is similar to studies of 
transitions for which the emissivity is proportional to $\rho^2$ that we discussed
earlier.}. However,
 in terms of the VCA this change is not important and we still expect to
see the emissivity index steepening as velocity slice thickness increases,
provided that velocity effects are present. In view of
this, results of Deshpande et al. (2000), who did not see such steepening,
can be interpreted as the evidence of the viscous suppression of
turbulence on the scales less than 1~pc. The fluctuations in this
case should be due to density and their shallow spectrum $\sim k^{-2.8}$ maybe related to damped magnetic structures below the viscous
cutoff (see Lazarian, Vishniac, \& Cho 2004).
This may be  also a consequence of the shallow 
density spectrum in compressible MHD 
(see Beresnyak, Lazarian \& Cho 2005). 

\begin{figure*}
\hbox{
 \includegraphics[height=.25\textheight]{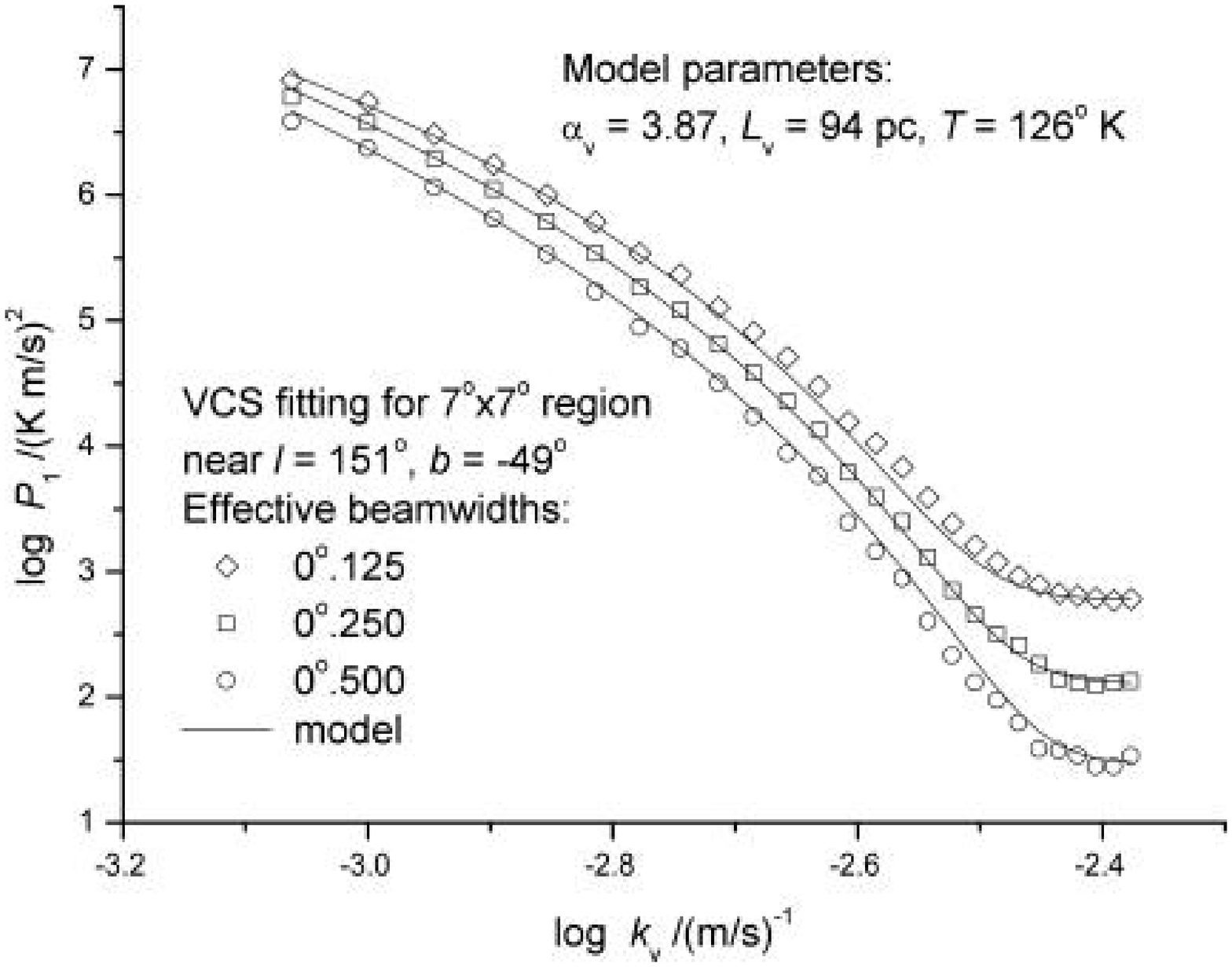}
 \includegraphics[height=.3\textheight]{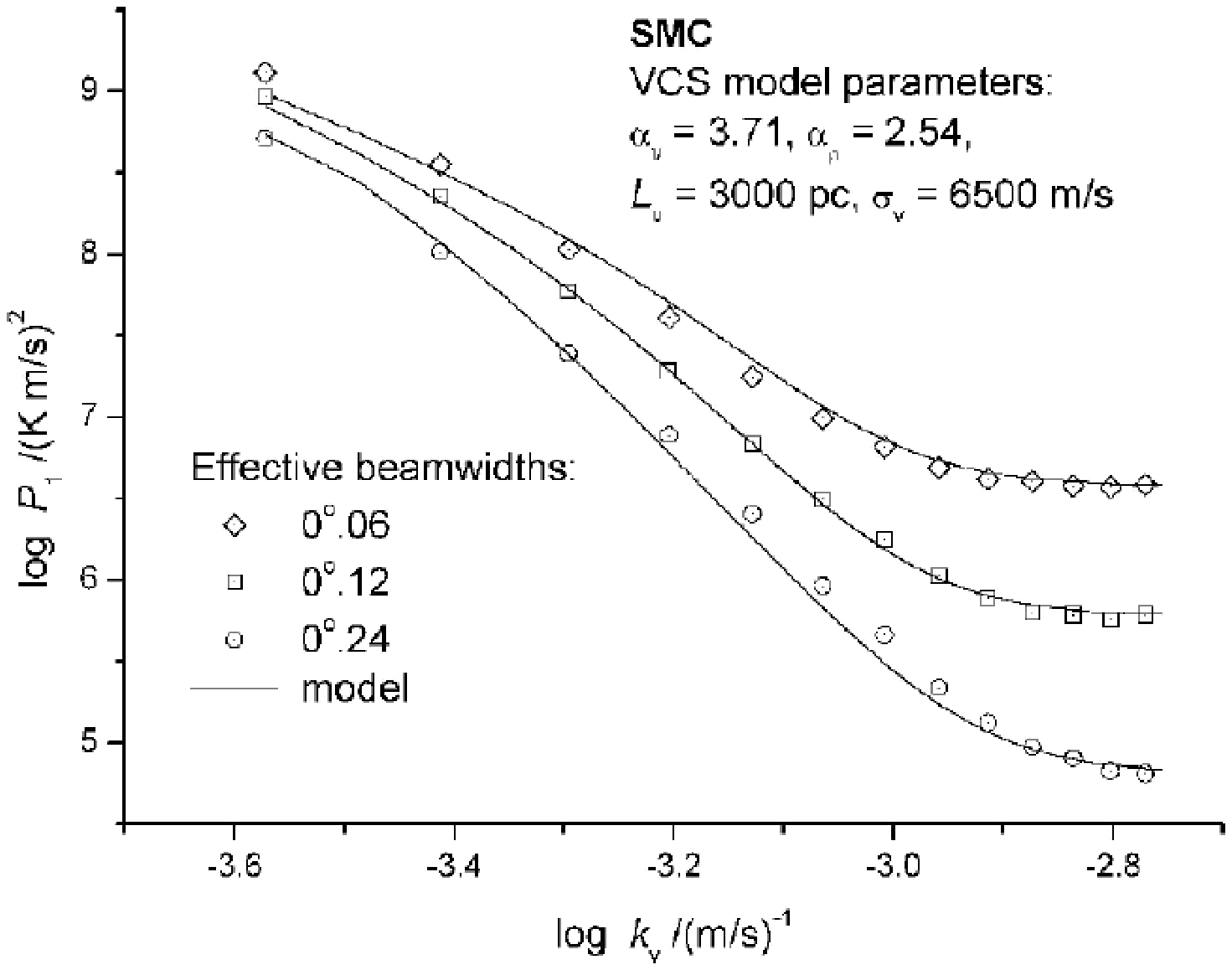}
}
  \caption{Fitting of turbulent models and observational data for different resolutions. {\it Left Panel}:
 Application of the VCS high latitude  HI Arecibo data. The spatial resolution of the maps was decreased to illustrate the VCS in both high and low resolution regimes. $\epsilon$ here is the power spectrum index, which for the Kolmogorov turbulence is $11/3$. The measured $\epsilon$ is in the range $[3.52; 3.57]$. The
 energy injection scale is 94 pc. {\it Right Panel}:
 Application of the VCS to Small Magellanic Cloud HI data. The spectral index for velocity is a bit more shallow. 
 The model corresponds to the energy injection at 3 kpc.From Chepurnov \& Lazarian.} 
 \label{galac}
\end{figure*}

Historically, the CO data was analyzed after integration over the entire emission line. 
Stutzki et al. (1998) presented the power spectra of  $^{12}$CO and $^{13}$CO 
fluctuations obtained via integrating the intensity over the entire emission line
for L1512 molecular cloud. 
Counter-intuitively, Stutzki et al. (1998) found for both isotopes the power spectrum
with a similar spectral index. According to LP04 this may correspond to optically thick
assymptotics (i.e. the integration range of velocities is larger than $V_c$ (see Eq.~{\ref{eq:abs_width2})).
 If the velocity fluctuations dominate, the expected index is universal
and equal to $-3$ (meaning $K^{-3}$),
if the density fluctuations dominate (see Table~2) the expected index is $-3+\gamma$
(meaning $K^{-3+\gamma}$).
The index measured in Stutzki et al. (1998) is $\sim 2.8$, which may either correspond to
$-3$ within the experimental errors, or more likely indicate that $\gamma\approx 0.2$,
i.e. the density spectrum is shallow. The latter possibility is indirectly supported by
$^{18}CO$ measurement for L1551 cloud in Swift (2006), used the VCA (observing
the changes of the channel map spectral index while changing the
velocity slice thickness) and obtained the shallow density spectrum with $\gamma \approx 0.2$,
while his measured velocity spectrum was approximately Kolmogorov (the index is $-3.72$).
Padoan et al. (2006) both successfully tested the VCA with high resolution numerical simulations
that included radiative transfer and applied the technique to Five College Radioastronomy Observatory
(FCRAO) survey of the Perseus molecular cloud complex. He obtained the velocity index around $-3.81$.



The practical applications of the VCA have only just began. Nevertheless, it has 
already provided some intriguing results
and proved to be a promising way of using the wealth of spectroscopic surveys for
studies of astrophysical turbulence. 

\subsection{Applying VCS}

Figure~\ref{galac} (left)  shows the results of our VCS-analysis of galactic high latitude data.
 Rather than  first correcting for the gas thermal broadening,
 then fitting the power law into the VCS spectrum as discussed in
 Lazarian \& Pogosyan (2006), we used our analytical expressions to 
 find the model that fits the data set corresponding to VCS for data
 at different spatial resolutions. The resolutions play for the VCS a similar\footnote{The important
 difference is that with VCS, we can restore the velocity spectrum for any resolution,
 while for the VCA, the velocity spectrum is available only for the thinnest slices.} role as the thickness
 of slices $\Delta v$ for the VCA and therefore PPV data cubes at different resolutions
 are non-trivially related, as far as the VCS analysis is concerned.  
 
 Fitting data to the models  opens ways of studying non-power law turbulence, e.g. turbulence 
at the injection or dissipation scales. It allows also studies of turbulence when thermal
broadening is important. Our results in Figure~\ref{galac} (left) show that the model of turbulence with spectrum steeper than Kolmogorov,
i.e. with $E_{v}\sim k^{-1.9}$,  the temperature\footnote{The analysis in LP06 shows that the
contribution of turbulence in warm gas to fluctuations in PPV is exponentially suppressed compared to that in cold gas.}  of gas around $130$K and a single injection
scale of $100$pc.

Different lines correspond to studies of maps with different resolution. 
The fitting  was done with a solenoidal component of turbulence only,
which results in variations of the fitting precision as the resolution changes\footnote{In general, both the solenoidal and potential
components of turbulence contribute to  the observed signal. The relative contribution of the two components varies with the
resolution. With higher precision data, one can potentially disentangle the two components by varying the resolution.}. This is the first example
of the analysis in which we obtained more than just the turbulence spectral index from the spectroscopic data. Our
first applications of a similar approach to  the VCA have also produced encouraging results. In particular, we improved the precision of
determining turbulence spectrum in HI of Small Magellanic Cloud (SMC) (see line 3 in Table~\ref{table}) and obtained the injection scale of around 4kpc,
and which corresponds to the ideas of exciting turbulence in SMC during its encounter with its
neighbor, Large Magellanic Cloud. Figure~\ref{galac} (right) illustrates the application of the VCS to the same set of SMC
data. The correspondence in the spectral indexes of velocity turbulence which were obtained by independent application of the VCA and VCS techniques is encouraging.

\subsection{Expected developments}

Our discussion above shows that fitting of the observed spectrum of fluctuations with turbulence models is more informative than just using asymptotic solutions. Although the theoretical work so far was aimed at obtaining the expected asymptotical analytical relations, the formalism developed in this process contains integral equations that can be used for the detailed fitting similar to
the one described in \S 9.2. We expect that both the VCA and VCS techniques will be used this way in future modeling. 

We also expect that VCA and VCS are to be used simultaneously, whenever possible. In addition, we expect that combining many emission and absorption lines will provide insight into astrophysical turbulence in its complexity. Indeed, as we discussed earlier,
the techniques are most sensitive to the colder component of gas. Therefore, for an idealized situation of two phase HI, the turbulence is being sampled mostly in cold phase. To sample turbulence in the warm phase or hot phase as, for instance, is the case for the clouds in the Local Bubble, one may use other species as tracers. Note, that the properties of turbulence may differ from one phase to another, which makes these type of studies very important.

\section{Alternative approaches}

\subsection{Velocity centroids}

As we mentioned in the introduction, a more traditional approach to turbulence studies includes 
 velocity centroids, i.e.
$S(\mathbf{X})=\int v_{z}\ \rho_{s}(\mathbf{X},v_{z})\ {\rm d}v_{z}$,
where $\rho_{s}$ is the density of emitters in the  PPV space.
Analytical expressions for structure functions\footnote{Expressions for the correlation 
functions are straightforwardly related to those of structure functions. 
The statistics of centroids using correlation functions was studied in a follow-up paper by 
Levier (2004).}   of centroids, i.e.
$\left\langle \left[S(\mathbf{X_{1}})-S(\mathbf{X_{2}})\right]^{2}\right\rangle $
 were derived in Lazarian \& Esquivel (2003). In that paper a necessary criterion
for centroids to reflect the statistics of velocity was established.
 Esquivel \& Lazarian (2005) confirmed the utility of the criterion
and revealed that for MHD turbulence
simulations it holds for subsonic or slightly supersonic
turbulence (see also Ossenkopf et al. 2006).
A subsequent study by Esquivel et al. (2007) showed that there are
fundamental problems with using centroids for supersonic turbulence.
This is in contrast to the VCA and the VCS  that provide reliable ways to study
supersonic turbulence.
At the same time, velocity centroids can be successfully used to study
anisotropies arising from the existence of the mean field (see \S12.1).

\subsection{Wavelets and Principal Component Analysis}

The use of different {\it wavelets} for the analysis of data is
frequently treated in the literature as different statistical
techniques of turbulence studies (Gill \& Henriksen 1990, Stutzki et al. 1998, Cambresy 1999, Khalil et al. 2006), which creates an
illusion of an excessive wealth of tools and approaches. In reality,
while Fourier transforms use harmonics of $e^{i{\bf kr}}$, wavelets use
more sophisticated basis functions, which may be more appropriate for
problems at hand.  In our studies we also use wavelets both to analyze
the results of computations (see Kowal \& Lazarian 2006) and synthetic
maps (Ossenkopff et al. 2006, Esquivel et al. 2007), along with or
instead of Fourier transforms or correlation functions. Wavelets may
reduce the noise arising from inhomgeneity of data, but we found in
the situations when correlation functions of centroids that we studied
were failing, a popular wavelet ($\Delta$-variance) was also failing
(cp. Esquivel \& Lazarian 2005, Ossenkopff et al. 2006, Esquivel et
al. 2007).  While in wavelets the basis functions are fixed, a more
sophisticated technique, Principal Component Analysis (PCA), chooses
basis functions that are, in some sense, the most descriptive.
Nevertheless, the empirical relations obtained with PCA for extracting
velocity statistics provide, according to Padoan et al. (2006), results for low order statistics, $(\delta v)^p$, where $p<0.5$ (see also Brunt et al. 2003), while the spectrum corresponds to $p=2$. In addition,
while the PI's research shows that for density spectra $E_{\rho}\sim
k^{-\alpha}$, for $\alpha<1$ both velocity and density fluctuations
influence the statistics of PPV cubes. 

It is also worrisome that no dependencies of PPV
statistics on density have been reported so far in PCA studies. We do know from the analysis
in LP00, that, for shallow density the fluctuations, PPV statistics should depend both on velocities 
and densities. Therefore no detection of the density spectrum
may reflect the problem of finding the underlying relations
empirically with data cubes of limited resolution. The latter may induce
a special kind of shot noise (Lazarian et al. 2001, Esquivel et al. 2003, Chepurnov \&
Lazarian (2006, 2008). One way or another, this seem to illustrate the difficulty of empirical
establishing of the relations between the PPV statistics and the underlying velocities and densities.

We feel that the analytical insight which was obtained in the process of VCA and VCS development should be used to get the
corresponding calibration of the techniques which use wavelet or PCA decomposition instead of the Fourier one. Such a calibration can be done in many cases analytically.

\subsection{Spectral Correlation Function}

{\it Spectral Correlation Function (SCF)} was developed by Alyssa Goodman's group at the 
same time  we developed the VCA technique
 (see Rosolowsky et al. 1999 and Lazarian 1999, respectively). Further
 development of the SCF technique in Padoan et al. (2001) removed the
 adjustable parameters from the original expression for the SCF and
 made the technique rather similar to VCA in terms of the
 observational data analysis. Indeed, both SCF and VCA measure
 correlations of intensity in PPV ``slices'' (channel maps with a
 given velocity window $\Delta v$) (see Figure~\ref{fig1}), but if SCF treats
 the outcome empirically, the analytical relations in Lazarian \&
 Pogosyan (2000) relate the VCA measures to the underlying velocity
 and density statistics.  In fact, we predicted several physically-motivated
 regimes for VCA studies. For instance, slices are ``thick'' for
 eddies with velocity ranges less than $\Delta v$ and ``thin''
 otherwise.  VCA relates the spectral index of intensity fluctuations
 within channel maps to the thickness of the velocity channel and to
 the underlying velocity and density in the emitting turbulent
 volume. We showed that much of the earlier confusion stemmed from
 different observational groups having used velocity channels of
 different thicknesses (compare, e.g.,Green 1993 and Stanimirovic et al. 1999). 
 In the VCA these variations of indexes with
 the thickness of PPV ``slice'' are used to disentangle velocity and
 density contributions. 
 
 We suspect that similar ``thick" and ``thin"
 slice regimes should be present in the SCF analysis of data, but they
 have not been reported yet.  While the VCA can be used for all the
 purposes the SCF is used (e.g. for an empirical comparisons of
 simulations and observations), the opposite is not true. In fact,
 Padoan et al. (2006) stressed that VCA eliminates errors inevitable
 for empirical attempts to calibrate PPV fluctuations in terms of
 underlying 3D velocity spectrum. With the VCA one can relate both
 observations and simulations to {\it turbulence theory}. Using explicit expressions
 for the VCA one can also study non-power law velocity fluctuations.

SCF was successfully used in Padoan et al. (2003) to relate the results of numerical simulations
and observations. The advantage of the VCA and VCS techniques is that they can also
relate the simulations and observations to the turbulence theory. Therefore, it looks advantageous
to repeat the analysis in Padoan et al. (2003) using both the VCA and VCS techniques.

\subsection{Identifying objects in PPV space}  

There exist a number of algorithms that identify shells and clumps using PPV data. The most widely-used
codes, namely "Gaussclumps" (Stutzki \& Guesten 1990) and "CLUMPFIND" (Williams, de Geus \& Blitz 1994)
have been used to identify clumps in many observational data sets. 

In view of our discussion of the complexity of the PPV space, one can potentially encounter problems arising
from the caustics created in the velocity space. Indeed, we know both from the analytical theory and numerical
testing that the ripples along the velocity axis are strongly affected by the turbulent velocities. This calls for
more studies aimed at understanding at what conditions the features identified in the PPV space are real and in
what cases they are caused by velocity crowding. Future research can identify the characteristic features of
the latter effect.

\section{Synergy and Future Work}

\begin{figure*}
\hbox{
 \includegraphics[height=.25\textheight]{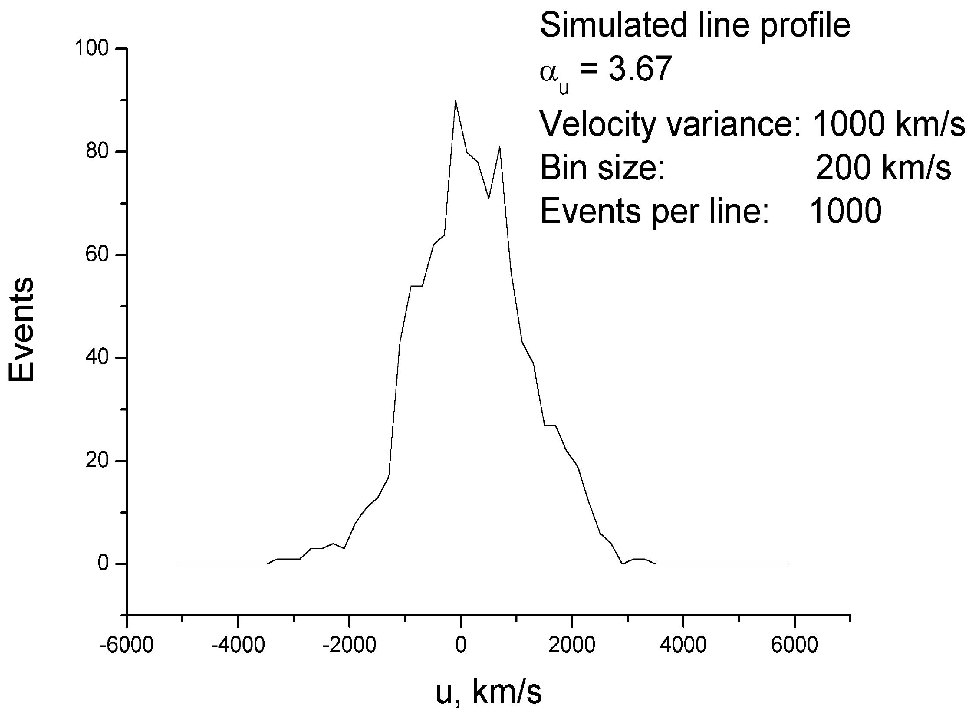}
 \includegraphics[height=.3\textheight]{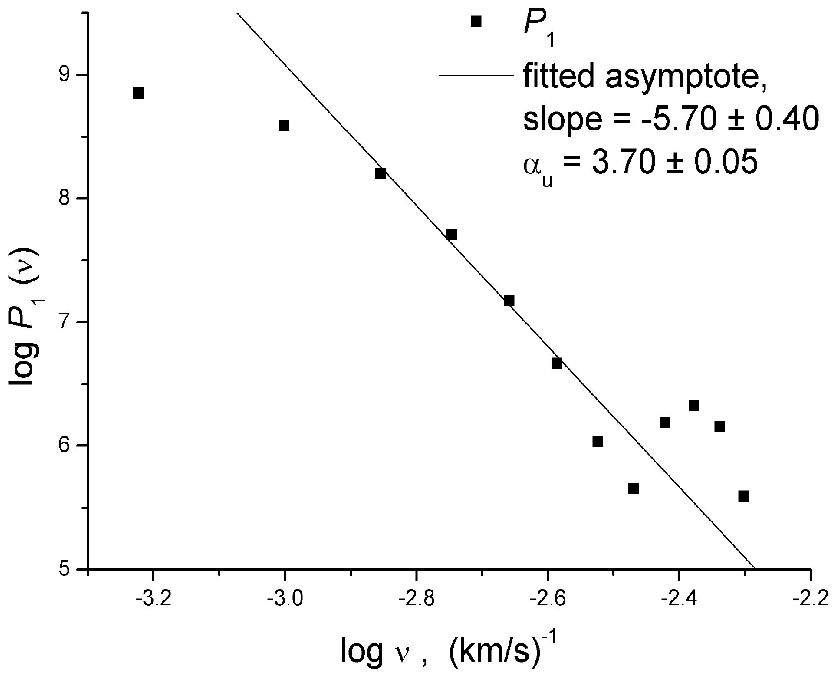}
}
  \caption{ Synthetic spectral line (left) and the velocity spectrum (right) that can be obtained with Constellation
  X for turbulence in clusters of galaxies. Calculations are done assuming one hour of observations.} 
  \label{X}
\end{figure*}

\subsection{Synergetic statistical studies}

We would like to stress, however, that both VCS and VCA are the ways of obtaining of 
the spectrum of turbulent velocities.The velocity spectrum is not the only characteristics that the
astronomers would like to know about the turbulence. 
Density spectrum is another
characteristics that is available from the techniques above. The spectra of magnetic fields are 
potentially available through measurements of Faradey rotation and synchrotron emission (see Cho \& Lazarian 2002, Haverkorn et al. 2008). However, even having 3 different spectra above, one does not have a full description of
turbulence. The problem is that the spectrum, as a measure, does not preserve the information about the
phases of turbulent motions. Below we list techniques that could, potentially, provide a complementary insight.

{\it Genus analysis} is a tool to study topology used in
Cosmology (Gott et al. 1990). The 2D ``genus'', $G(\nu)$, is the
difference between the number of regions with a projected density
higher than $\nu$ and those with densities lower than $\nu$.  Unlike
visual inspection, the genus quantifies the topology and allows us to
compare numerical results with observations. In Kowal et al. (2007) we
performed a systematic study of genus statistics for synthetic maps obtained via MHD turbulence
simulations and studied the variations of genus with sonic and Alfven Mach numbers. We
determined the cases when these variations were more prominent than
the corresponding variations in spectral index. Genus was applied to
observational data in Lazarian et al. (2002), Kim \& Park (2007), Chepurnov et al. (2008).  
 Note that genus at large scales (obtained with smoothing of
the high resolution maps) is expected to be different for models of
ISM where hot hydrogen forms tunnels (see Cox 2005) and in the
classical McKee \& Ostriker (1977) model.

{\it Bispectrum} is another tool which was first used in 
Cosmology (see Scoccimarro 2000 for a review). Its prospects for ISM
studies were discussed in Lazarian (1999) and Lazarian, Kowal \& Beresnyak (2008).
 The standard power spectrum
is obtained by multiplying the Fourier transform of a given ${\bf k}$
by its complex conjugate. In this process, the complex component is
eliminated along with all information about {\it wave phases}. In the
Bispectrum technique, the correlation of the Fourier transforms of
different ${\bf k}$'s is studied, and the wave phases are kept,
allowing the turbulent fields corresponding to the same spectrum to be
distinguished by their bispectrum. This method is ideal for the study
of scale-correlations in multidimensional problems. Recently
we applied  bispectrum to the synthetic
column density maps obtained from numerical simulation cubes. We showed
that MHD turbulence presents a larger band of non-linearly
interacting waves than non-magnetized turbulence does (Burkhart et al. 2008). 

{\it Studies of Anisotropy}. A technique for studies of magnetic
field direction, which makes use of preferred correlation of turbulent
motions along the field, was proposed in Lazarian et al. (2002). The
technique was originally discussed in terms of channel maps and velocity
centroids (see also Vestuto et al. 2002) The
application of it to synthetic maps in Esquivel \& Lazarian (2005)
showed that the anisotropies in the statistics of velocity centroids
reflect well the projected direction of the magnetic field, even for
high Mach number turbulence. This makes the technique very appealing
to studies of magnetic fields in molecular clouds. A study of 
the anisotropy using the PCA decomposition was successfully performed 
by (Heyer et al. 2008).

\subsection{Prospects of quantiative studies of PPV data}

VCA and VCS are two new techniques that emerged as an attempt to disentangle velocity and
density contributions in studies of turbulence with spectroscopic data. They evolved from modest
attempts to explain the existing velocity channel data in LP00 to formulating new ways of data
analysis as in LP06, LP08 and Chepurnov \& Lazarian (2008). 

In LP00, LP04, LP06 we used HI as an example of species to which the technique
is to be applied. 
Using heavier species that have lower thermal Doppler width of spectral lines
allows one to study turbulence up to smaller scales.  In addition, we would
like to stress that the VCS technique can be used at different wavelength. For
instance, the X-ray spectrometers with high spatial resolution can be used
to study of turbulence in hot plasma. In particular, the potential of VCS is
high for studies of turbulence in clusters of galaxies (cf.
Sunyaev et al. 2003 and references therein).  A simulated
example of such a study with the future mission Constellation X is provided in Figure~\ref{X}. 

Studies of turbulence in objects which are poorly resolved spatially is a
natural avenue for the VCS applications. Interestingly enough, in this case
one can combine the absorption line studies, which would provide the 
VCS for the pencil beam, i.e. for the high resolution, with the emission 
studies that would provide the VCS in the poor resolution limit. Potentially,
both velocity and density spectra can be obtained this way.

The importance of this work goes beyond the actual recovery of the
particular power-law indexes. First of all, as we mentioned earlier, 
the techniques can be generalized
to solve the inverse problem to recover non-power law turbulence spectra.
This may be important for studying turbulence at scales at which
either injection or dissipation of energy happens. Such studies are important
for identifying astrophysical sources and sinks of turbulent energy.
Second, studies of the VCS transition from low resolution to high resolution
regimes (see Figure~\ref{galac}) allows one to separate thermal and non-thermal contributions
to the line-widths as it is discussed in LP06. This could both
test the thermal correction that can be applied to extend
the power-law into sub-thermal velocity range (see also Chepurnov
\& Lazarian 2006b) and enable studies of temperature distribution
of the gas in atomic clouds (cf. Heiles \& Troland, 2003). 

In \S 6 we mentioned several advantages of using VCA and VCS simultaneously. One may mention an additional
one. In compressible fluid the power spectrum of velocity can be decomposed into
the spectra of solenoidal (incompressible) and potential (compressible) motions. Interestingly enough, the VCA and VCS depend
on these two components differently. This opens prospects of studying the effects of compressibility by combining the two techniques. 

We feel that one should not feel constrained by the framework of the VCA and VCS techniques. The most important, in terms of
theory, was the development of  the general
description of the correlations within PPV cubes for different situations, including self-absorbing data and saturated absorption lines. This opens avenues to developing completely new ways of analyzing 
observational data. As we discussed in \S 10.2 the use of wavelets instead of Fourier transforms may be a straightforward
generalization of the techniques. More sophisticated work may be required to get higher order correlations and bispectrum from the PPV data. 

\subsection{Outlook onto Big Picture}

 \begin{figure}
   \begin{center}
 \includegraphics[width=.5\textwidth]{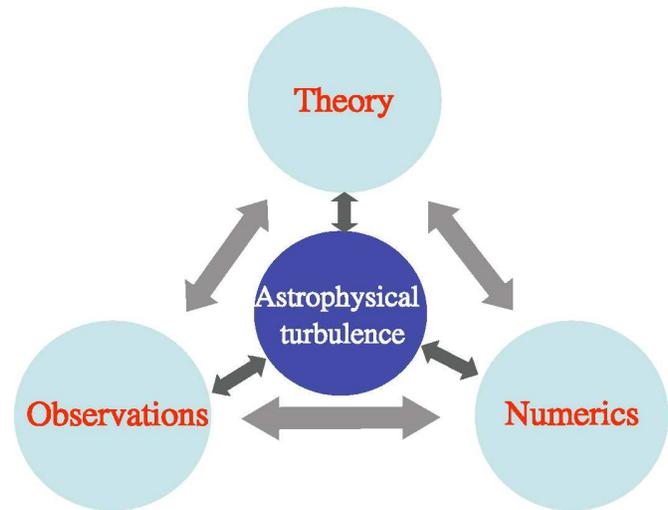}
 \end{center}
  \caption{Big Picture: interrelation between the approaches to studying astrophysical turbulence.}
\label{alex}
\end{figure}

Several decades ago measurements of spectral profiles were rather challenging. The situation is
radically different these days where the challenge is to use productively the enormous wealth of spectroscopic surveys. Therefore
the development of the theoretical description of the PPV statistics, which resulted in the birth of the VCA and VCS techniques is a
very timely achievement. 

Astrophysical turbulence is a very tough subject and dealing with it requires a cohesive use of analytical, numerical and observational approaches (see Figure \ref{alex}). Analytical studies most often predict power spectra for turbulence in different 
regimes (see Galtier et al. 2002, Goldreich \& Sridhar 1995, Lazarian et al. 2004, Boldyrev 2006, Lithwick et al. 2007, Beresnyak \& Lazarian 2008, Chandran 2008). At relatively high effective viscosities and resistivities, which are the only ones that are available
with the present-day computational facilities, these predictions can be tested (see Cho \& Vishniac 2000, Maron \& Goldreich 2001,
Cho, Lazarian \& Vishniac 2002ab, 2003, Cho \& Lazarian 2002, 2003, Beresnyak \& Lazarian 2006, 2008, Galtier 2008). We expect the VCS and VCA techniques to test the turbulence spectra in astrophysical circumstances.

We expect that as the turbulence theory matures, it starts posing more sophisticated questions, e.g. on the scaling of the higher
order velocity correlations (She \& Leveque 1994, Dubrulle 1994, see Biskamp 2003). In view of this the statistical description of the PPV space obtained so far can serve as a starting ground for developing new techniques. 

This is not the only role of the quantitative description of the PPV statistics, however. Astrophysical turbulence happens in complex environments and has multiple sources and sinks\footnote{The energy is being injected by supernovae explosions, stellar winds, magnetorotational instability etc. In addition, cosmic rays may redistribute the turbulent energy absorbing it
at the large scales and reinjecting at small spatial scales (Lazarian \& Beresnyak 2006).} . Many of numerical simulations, for instance, numerical simulations of interstellar medium, try to simulate the astrophysical settings in all its complexity. Theory in these situations provides rough guidance and insight, while through observational studies of actual shape of the power spectra can clarify many of the relevant issues. 

For all the applications it is important that the developed techniques can be used for studying turbulence in different phases of the astrophysical media. Indeed, the VCA and VCS techniques can employ various spectral lines: emission and absorption, optically thin and thick, saturated and not saturated.


\begin{acknowledgements}
I thank Alexey Chepurnov, Dmitry Pogosyan for their input.  The research is supported by
by the NSF Center for Magnetic Self Organization in
Laboratory and Astrophysical Plasmas and NSF grant AST 0808118. Helpful comments by the referee were appreciated. I thank Blakesley Burkhard for
reading the manuscript and her suggestions. 
\end{acknowledgements}


\end{document}